\documentclass[acmlarge]{acmart}

\makeatletter

\def\@ACM@copyright@check@cc{}

\newcount\ACM@corresponding@author
\newcount\ACM@address@author

\def\@mkauthorsaddresses{%
  \ifnum\num@authors>1\relax
    Authors' %
  \else
    Author's %
  \fi
  Contact Information:\space
  \bgroup
  \ACM@address@author=0\relax
  \def\streetaddress##1{%
    \ClassWarning{\@classname}{%
      ACM no longer collects authors' postal addresses.
      I am ignoring your street address}%
    \unskip\ignorespaces
  }%
  \def\postcode##1{%
    \ClassWarning{\@classname}{%
      ACM no longer collects authors' postal codes.
      I am ignoring your postal code}%
    \unskip\ignorespaces
  }%
  \def\position##1{\unskip\ignorespaces}%
  \gdef\@ACM@institution@separator{, }%
  \def\institution##1{%
    \unskip
    \@ACM@institution@separator
    ##1%
    \gdef\@ACM@institution@separator{ and }%
  }%
  \def\city##1{\unskip, ##1}%
  \def\state##1{\unskip, ##1}%
  \renewcommand\department[2][0]{%
    \unskip\@addpunct, ##2%
  }%
  \def\country##1{\unskip, ##1}%
  \def\and{%
    \unskip;
    \gdef\@ACM@institution@separator{, }%
  }%
  \def\@author##1{%
    \advance\ACM@address@author by 1\relax
    ##1%
    \ifnum\ACM@address@author=\ACM@corresponding@author
      \space{\normalcolor(corresponding author)}%
    \fi
  }%
  \def\email##1##2{%
    \unskip, \nolinkurl{##2}%
  }%
  \addresses
  \egroup
}

\makeatother

\usepackage{multirow}
\usepackage{subcaption}
\usepackage{makecell}
\usepackage{enumitem} 
\usepackage{xspace}
\usepackage[normalem]{ulem}
\usepackage{adjustbox}
\usepackage{pifont}
\usepackage{tabularx}

\useunder{\uline}{\ul}{}

\newcommand{\cmark}{\textcolor{green}{\ding{51}}} 
\newcommand{\xmark}{\textcolor{red}{\ding{55}}}   

\newcommand{\REV}[1]{{#1}}
\newcommand{\TOPK}[1]{{#1}}
\newcommand{\MINOR}[1]{{#1}}
\newcommand{\LONG}[1]{{#1}}
\newcommand{\COMPATIBLE}[1]{{#1}}
\newcommand{\REPRESENT}[1]{{#1}}
\newcommand{\INTERVIEW}[1]{{#1}}
\newcommand{\FINALREV}[1]{{#1}}

\newcommand{\done}[1]{}

\newcommand{\add}[1]{#1}
\newcommand{\deleted}[1]{}

\newcommand\sys{\texttt{AID}\xspace}

\newcommand\ipvact{IPI-related actions\xspace} 

\newcommand{\smallpm}[2]{#1{\scriptsize$\pm$#2}}

\AtBeginDocument{%
  \providecommand\BibTeX{{%
    \normalfont B\kern-0.5em{\scshape i\kern-0.25em b}\kern-0.8em\TeX}}}

\setcopyright{cc}
\setcctype{by}
\acmJournal{IMWUT}
\acmYear{2026} \acmVolume{10} \acmNumber{3} \acmArticle{183}
\acmMonth{9} \acmDOI{10.1145/3831660}

\begin{document}


\title{\REV{Towards On-Device Evidence Gathering for Intimate Partner Infiltration: A Feasibility Study for Joint Identity–Action Detection}}

\author{Weisi Yang}
\affiliation{%
  \institution{Northwestern University}
  \city{Evanston}
  \state{IL}
  \country{USA}}
\email{weisi@u.northwestern.edu}
\orcid{0009-0006-8566-5475}

\author{Shinan Liu}
\affiliation{%
  \institution{University of Hong Kong}
  \city{Hong Kong}
  \country{China}}
\email{shinan6@hku.hk}
\orcid{0000-0002-6170-2167}

\author{Feng Xiao}
\affiliation{%
  \institution{Georgia Institute of Technology}
  \city{Atlanta}
  \state{GA}
  \country{USA}}
\email{fengxiao3583@gmail.com}
\orcid{0009-0008-6755-2973}

\author{Nick Feamster}
\affiliation{%
  \institution{University of Chicago}
  \city{Chicago}
  \state{IL}
  \country{USA}}
\email{feamster@uchicago.edu}
\orcid{0000-0001-9315-5201}

\author{Stephen Xia}
\affiliation{%
  \institution{Northwestern University}
  \city{Evanston}
  \state{IL}
  \country{USA}}
\email{stephen.xia@northwestern.edu}
\orcid{0000-0001-5713-8885}

\renewcommand{\shortauthors}{Yang et al.}

\begin{abstract}
    \REV{Intimate Partner Infiltration (IPI) refers to phone-side privacy infiltration in intimate or close relationships, often enabled by physical access to a person's smartphone and discussed in technology-facilitated Intimate Partner Violence (IPV) contexts.} Unlike conventional cyberattackers, IPI perpetrators leverage proximity and personal knowledge to circumvent standard protection, underscoring the need for targeted interventions\REV{, motivating device-side tools that surface such risk evidence for later review}. While prior works have extensively studied IPV, and some have provided tailored and effective solutions such as security clinics, they are \REV{necessarily episodic and human-expert-intensive, and offer limited automated visibility into what happens on a smartphone between support sessions}. \INTERVIEW{Guided by a formative interview with experts (n=5)}, we take the first exploration \REV{into gathering IPI-risk evidence} from a mobile system perspective and present \sys, \underline{A}utomated \underline{I}PI \underline{D}etection, a data-driven system that continuously logs unauthorized access and suspicious behaviors on smartphones. \TOPK{In a controlled 27-participant study, \sys achieves an end-to-end F1 score of 0.928 with a 7.0\% false positive rate for Top-1 phone-side risk flagging; when preserving top-3 candidate action categories as report context, \sys achieves an F1 score of 0.981 and a false positive rate of 1.6\%}. These findings demonstrate \sys's potential as an \REV{evidence-support tool that complements current clinic-based interpretation and safety-planning}.

\end{abstract}



\maketitle


\section{INTRODUCTION}

Intimate Partner Violence (IPV) is a prevalent issue around the world. In the United States alone, 47\% of women and 44\% of men are affected by different forms of IPV during their lifetime~\cite{leemis2022nisvs}. Perpetrators of IPV often monitor, control, and harass victims through technology and physical devices that are prolific in our daily lives~\cite{ceccio2023sneaky,rogers2023technology,freed2018stalker,bellini2023digital}. For example, an IPV abuser might exploit a home router to monitor a victim’s smartphone~\cite{tseng2020tools} and internet activity~\cite{freed2018stalker}, a GPS-tracker to give the abuser information about the victim’s real-time location, or a hidden camera to spy on the victim’s daily activities~\cite{ceccio2023sneaky}. \REV{Among these technology-facilitated tactics, this work focuses on a narrower phone-side pattern that we call Intimate Partner Infiltration (IPI). In cases of IPI, an intimate partner or otherwise close person gains physical or credentialed access to a victim’s smartphone and performs privacy-sensitive actions on the device}. For example, once \REV{a close person} gains access to the victim’s phone, s/he may change passwords, delete important emails, or snoop through personal messages to gain ammunition for harassing the victim. While there has been extensive research in IPV/IPI across psychology, health, engineering, and computer science, the current solutions for mitigating IPI are limited.

Recently, a class of approaches called clinical computer security has gained traction for supporting people affected by technology-facilitated IPV and IPI-related harms~\cite{havron2019clinical,tseng2022care}. These approaches are administered through \textit{Security Clinics}~\cite{bellini2024abusive,tseng2022care,havron2019clinical} and leverage a combination of clinical interviews, consultations, and professional support to provide tailored guidance. These human-centered approaches are essential, but they are necessarily episodic and expert-intensive. For instance, logistical barriers and the limited supply of security experts make it more difficult for residents of rural areas to access these services. \REV{Additionally, reconstructing past device activity through self-report can be incomplete or emotionally taxing~\cite{ramjit2024navigating}, motivating complementary device-side evidence that can be reviewed with trained support professionals.}

In this work, we explore the possibility of continuously detecting and classifying phone-side IPI-risk events locally. \REV{An IPI-risk event is formulated as the conjunction of two observable components: non-owner-like device use and an IPI-relevant phone-side action}. In doing so, security clinics can review timestamped reports about detected non-owner-like sessions and ranked IPI-relevant action categories at all times of day, even outside clinical sessions. Such a digital tool could lead to more individualized and finer-grained monitoring of IPI events across a larger population of victims. To the best of our knowledge, \REV{\textbf{this is the first work that takes a systems-based approach for detecting and flagging IPI-risk evidence on-device}}.

While there are several seemingly related fields that have been extensively explored, such as anomaly detection, continuous authentication, and human activity recognition (HAR), there are several unique challenges that make them fall short in the IPI context. A summary is provided in Table~\ref{tab:related-landscape} and discussed next.

\noindent
\textbf{1. IPI exploits trust rather than technical vulnerabilities.} Rather than exploiting technical vulnerabilities, an intimate adversary typically leverages physical proximity, shared access, or familiarity with the device owner (e.g., knowledge of the owner’s routines or authentication credentials) instead of technical exploits. For instance, intimate partners are often already registered on shared devices (e.g., fingerprints previously added to the device’s authentication system), or can obtain access through educated guessing and coercion. This reduces the effectiveness of traditional security measures, such as biometrics or passcodes, and allows the abuser to bypass the need for sophisticated hacking. The unique combination of \textbf{physical proximity} and \textbf{intimate knowledge of the victim’s behaviors} allows anyone in an intimate relationship to become a potential abuser, regardless of their technical background.

\noindent
\textbf{2. Identity-only detection is insufficient; IPI-risk detection requires joint identity--action reasoning.} A logical first solution for detecting IPI is to simply detect whether someone other than the owner is using the phone (e.g., continuous authentication and anomaly detection) through built-in sensors. However, not all activities from a non-owner have malicious intent. For example, it is common for someone to borrow another phone to watch videos. On the other hand, simply classifying behaviors and activities (e.g., HAR and behavior studies) is also not sufficient. For example, changing passwords on a phone is a fairly standard procedure if the phone’s owner initiates, but may be a sign of abuse if the partner initiates. Existing bodies of work generally focus on detecting either \textit{identity} or \textit{event}, leading to high false positives, but monitoring IPI requires both.

\noindent
\textbf{3. Detecting user action is not standard activity recognition.} Most smartphone-based HAR research generally focuses on detecting activities with \textit{physical movements} (e.g., walking, gestures, etc.), where the phone is somewhere moving with the user’s body, rather than the user directly interacting with the phone itself. We show in our evaluation that leveraging such techniques and sensors performs poorly on detecting IPI behaviors, where the phone is generally moving much less and the user is directly interfacing with the phone (e.g., typing).

\noindent
\textbf{4. Stealthy constraints.} The system must operate stealthily, avoiding detection by abusers. If an abuser becomes aware of its existence, the victim could face escalated risks of harm. This challenge limits the sensors, input signals, and implementation methods that we can leverage, as the system cannot induce any visible indicator to the user. As such, many of the common sensing modalities used in related fields, such as HAR and anomaly detection, are not suitable for use in the IPI context.

\noindent
\textbf{Proposed work.} Building on prior human-centered IPV research and findings, we take, to the best of our knowledge, the first step toward detecting phone-side IPI-risk events from a systems-driven approach, leveraging device-internal sensors and signals. First, we formalize a novel threat model for IPI, including constructing a taxonomy of IPI behaviors reported in literature. Next, we propose an Android-based prototype \sys, \underline{A}utomated \underline{I}PI \underline{D}etection system, for continuous detection of potential IPI behaviors based on our taxonomy and threat model. \sys leverages streams of multimodal sensing data and a unified dual-branch architecture to detect IPI events by jointly analyzing 1) non-owner access and 2) behavioral actions that are indicative of IPI. We carefully design \sys for low visibility to users of the device by using a deceptive UI and constraining its access to only data streams that can be collected and processed in the background. \sys is also privacy preserving by performing inference and keeping private user information locally. \TOPK{Through a 27-person controlled feasibility study, \sys achieves an F1 score of 0.928 with a 7.0\% FPR under the primary Top-1 end-to-end candidate IPI-risk event flagging task. When the report retains the top-3 candidate action categories as contextual evidence for later review, \sys reaches an F1 score of 0.981 with a 1.6\% FPR}.

\textbf{We envision \sys as an evidence gathering tool for automated proactive detection of IPI-risk events on mobile devices, which potentially benefits users and current security clinics by providing an additional source of objective and informative evidence}. To summarize, our main contributions are:

\begin{itemize}[leftmargin=*,label=\raisebox{0.5ex}{\scalebox{1.2}{$\bullet$}}]
\item We propose and formalize a novel threat model for physical Intimate Partner Infiltration of smartphones. Our model jointly analyzes \textit{who} holds the phone, \textit{what} they do, and provides a taxonomy of IPI tactics distilled from prior IPV literature, revealing a concrete target and scheme for automated detection. (Section~\ref{sec-3-threatModel}).

\item We design \sys (\underline{A}utomated \underline{I}PI \underline{D}etection) for continuously assessing IPI-relevant risks on smartphones. \sys employs a unified dual-branch architecture that analyzes non-owner phone usage and fine-grained behavioral patterns as formalized by our threat model. To strengthen its privacy preservation and safety, \sys processes data locally and uses only zero-permission streams to minimize user-facing perceptual cues. (Sections \ref{sec-4-sysdesign}--\ref{sec:stealthy_functioning}).

\item \TOPK{Through a 27-person controlled feasibility study, \sys~achieves 0.928 F1 with a 7.0\% FPR under the primary Top-1 end-to-end candidate IPI-risk event flagging task. When retaining the top-3 candidate behaviors, \sys reaches 0.981 F1 with a 1.6\% FPR, showing that it can preserve useful ranked evidence for expert interpretation (Section~\ref{sec-5-eval})}.

\end{itemize}

\section{BACKGROUND AND MOTIVATION}\label{sec-2-background}

In this section, we discuss two categories of work related to IPI: (1) human-centered approaches, such as security clinics developed for IPV victims, and (2) technical solutions drawn from adjacent fields.

\begin{table}[t!]
\small
\centering
\caption{Landscape of related technical method spaces versus \sys. Symbols: \cmark = satisfies, \textcolor{orange}{$\sim$} = partially, \xmark~= does not satisfy. 
}
\label{tab:related-landscape}
\begin{adjustbox}{width=\textwidth}
\begin{tabular}{@{}lllll|l@{}}
\toprule
\shortstack{\textbf{Characteristics}\\~\\~\\~} &
\shortstack{\textbf{\shortstack{Anomaly\\Detection}}\\ \small(e.g.\,\cite{DBLP:conf/ndss/MirskyDES18,phillip2023argus})} & 
\shortstack{\textbf{\shortstack{Continuous\\Authentication}}\\ \small(e.g.\,\cite{fereidooni2023authentisense,xu2014towards})} & 
\shortstack{\textbf{\shortstack{Human Activity\\Recognition}}\\ \small(e.g.\,\cite{zhang2022deep,liu2023amir})} &
\shortstack{\textbf{\shortstack{Digital Behavior\\Study}}\\ \small(e.g.\,\cite{OTEBOLAKU201633,xu2024autolife})} &
\shortstack{\textbf{\sys\ (Ours)}\\~\\~} \\
\midrule

Purpose &
Detect cyber intrusions &
Identity verification &

Physical activity recognition &
Large-scale context sensing &
\REV{Detect IPI-risk events} \\

Identity Check &
-- &
\cmark &

\xmark &
\xmark &
\cmark \\

Action Check &
-- &
\xmark &

\xmark &
\textcolor{orange}{$\sim$} &
\cmark~IPI-tailored actions (Section \ref{subsec:ipi_taxonomy})\\

Privacy cost &
\textcolor{orange}{$\sim$} &
\textcolor{orange}{$\sim$} &

\textcolor{orange}{$\sim$} &
\xmark &
\cmark \\

Output &
\textcolor{orange}{$\sim$}~Anomaly probability &
\textcolor{orange}{$\sim$}~Owner probability &

\xmark~Activity probability &
\textcolor{orange}{$\sim$}~Activity probability / diary &
\cmark~$\langle$IPI risk, identity, action$\rangle$ \\

Threat model &
\textcolor{orange}{$\sim$}~Primarily external &
\textcolor{orange}{$\sim$}~Outsider / casual insider &

-- &
-- &
\cmark~Credentialed insider (IPI abuser) \\

Stealthiness &
\xmark &
\xmark &

-- &
-- &
\cmark \\

Performance (F1 / FPR) &
0.540 / 0.276  & 0.908 / 0.181 & 
0.661 / 0.780 & -- &
\textbf{\TOPK{0.928 / 0.070 (Top-1); 0.981 / 0.016 (Top-3)}} \\
\bottomrule
\end{tabular}
\end{adjustbox}
\end{table}

\subsection{IPV-Related Research}

Prior research on technology-facilitated IPV has mainly provided qualitative insights into victims’ experiences and perceptions of how perpetrators misuse technology, such as how abusers deploy GPS and audio surveillance tools~\cite{freed2017digital,freed2018stalker} or leverage online forums for Intimate Partner Surveillance (IPS) and other technology-driven abuses~\cite{bellini2023digital,tseng2020tools}. Other research has also emphasized the need for legal frameworks and defensive tools to combat IPV and address the widespread availability of spy devices~\cite{thomas2021sok,ceccio2023sneaky}. A range of mitigation solutions has emerged in response to these findings, which we detail next.

\noindent
\textbf{Security clinics.} Security clinics provide in-person consultations for victims, offering tailored support to identify risks and rebuild a sense of safety~\cite{havron2019clinical,tseng2022care}. \add{Guided by this, some clinics have been established with proven efficacy\deleted{~\cite{ceta,mtc,cltc,cuomo2022tecc}}\add{, such as Clinics to End Tech Abuse (CETA) in New York City~\cite{ceta}, Madison Tech Clinic (MTC) in Madison~\cite{mtc}, the UC Berkeley Center for Long-Term Security (CLTC)~\cite{cltc}, and Seattle-based Technology-Enabled Coercive Control (TECC) Clinic~\cite{cuomo2022tecc}}.} While these approaches have demonstrated significant promise, \deleted{scaling their services beyond local communities presents considerable challenges}\add{challenges appear with the following bottlenecks}. 

\add{First, the clinic resources for each IPI-affected individual are constrained. This includes the time slots available to clients for clinical therapy and consultations, the high cost to train and involve highly skilled experts and staff for customized services, and the geographic barriers that limit rural residents' access. All of them pose barriers for clinics to scale. Second, security clinics usually proceed via conversations between experts and clients, requiring self-reporting which could potentially be incomplete due to safety constraints, recall burden, or emotional distress~\cite{ramjit2024navigating} for the survivor to recount.} \REV{These limitations motivate complementary tools that can organize phone-side traces into structured evidence candidates for later expert review and assessment.}

\deleted{One \textit{major limitation} is their specific physical location, which restricts access for victims outside those areas. Furthermore, these clinics require the involvement of technical experts who possess specialized knowledge to address the complex and evolving nature of technology-facilitated abuse. Recruiting and training such experts, as well as ensuring ongoing support, is resource-intensive and difficult to sustain at scale. This combination of geographic constraints and reliance on highly skilled personnel highlights the barriers to replicating models like the Clinics to End Tech Abuse (CETA) in New York City~\cite{ceta}, Madison Tech Clinic (MTC) in Madison~\cite{mtc}, the UC Berkeley Center for Long-Term Security (CLTC)~\cite{cltc}, and Seattle-based Technology-Enabled Coercive Control (TECC) Clinic~\cite{cuomo2022tecc} in broader contexts. These challenges underscore the need for scalable, accessible, and resource-efficient digital approach to combat IPI on a wider scale.}

\noindent\textbf{Digital tools.} Although basic tools exist to help users remove browser histories and delete digital traces to safeguard their privacy~\cite{arief2014sensible,havron2019clinical}, or provide education and safety-planning resources~\cite{myplanapp}, \REV{they do not directly support structured review of phone-side IPI-risk actions. For example, existing tools provide limited support for organizing evidence candidates around non-owner access and selected phone-side actions such as impersonating the victim, viewing private content, modifying account settings, or changing device/app data.}

\subsection{Related Technical Domains and Their Limitations in IPI Context}
\label{subsection:related_technical_domains}

Existing evidence-gathering heavily relies on human-intensive approaches. In contrast, we aim for an automated solution to address this problem. Although several well-established domains---such as anomaly detection, user authentication, and human activity recognition (HAR)---appear relevant, our empirical results (Section \ref{subection:e2eeval}) reveal significant shortcomings: high false positive rates of 27.6\%, 18.1\%, and 78.0\%, respectively. These levels of inaccuracy are deeply risky for potentially creating unnecessary concerns. As summarized in Table \ref{tab:related-landscape}, a core reason for these failures is that existing methods do not \emph{jointly} reason about \emph{who} is using the phone (identity) and \emph{what} they are doing (action).  The remainder of this subsection explains each domain’s limitations in the context of IPI; Section \ref{subsec:formulating} then introduces our joint identity–action formulation that addresses part of this gap.

\noindent
\textbf{Anomaly detection} produces binary outputs indicating whether anomalies occur in the input data sequence. Prior works have explored anomaly detection in domains such as network intrusion \cite{DBLP:conf/ndss/MirskyDES18,robertson2010effective} and  IoT security \cite{sikder20176thsense,phillip2023argus}. 
However, detecting \ipvact involves additional human-centered considerations. \textit{First}, anomaly detection systems often rely on Personally Identifiable Information (PII), such as user logs and network traffic, raising privacy concerns. \textit{Second}, conventional anomaly detection solutions for computational systems can notify an operator as soon as the detection confidence passes a threshold. Such immediate alerting is not suitable for IPI evidence gathering because it could lead to the risk of discovery, confrontation, or escalation. \textit{In addition}, many \ipvact resemble normal usage patterns, undermining the core assumption in anomaly detection that malicious events deviate sharply from benign or risk-neutral events \cite{chandola2009anomaly,malaiya2019empirical}.

\noindent
\textbf{User authentication}: Smartphone user authentication falls into two categories: \textit{one-time} and \textit{continuous} methods. While one-time verification (e.g., PINs, facial recognition) initially prevents casual intruders, it is insufficient in IPV settings where abusers may know these credentials~\cite{tseng2020tools,xu2014towards}. Continuous authentication, drawing on biometric or behavioral data (e.g., motion sensors~\cite{fereidooni2023authentisense,centeno2018mobile,liu2023amir}, touch interactions~\cite{frank2012touchalytics,xu2014towards}, or hybrid approaches~\cite{deb2019actions,acien2019multilock}), offers stronger security by persistently checking whether the current user matches the legitimate owner. 

On the surface, user authentication is well-suited to this problem as it can flag non-owner users of the phone. However, a critical nuance is that the ``intruder'' may not always pose harm. \textit{Benign sharing} -- such as lending a phone to a friend or family member -- is common in close relationships, yet existing systems 
\textit{are primarily designed to detect unauthorized access by strangers, without considering the user's \REV{interactions} on the device.} This mismatch can cause benign interactions to be misclassified as intrusions, leading to unnecessary concern or false alarms. Besides, some authentication methods rely on IPI-sensitive data such as location \cite{deb2019actions} and touch coordinates \cite{cai2024famos}, which could expose private information like location traces or typed passwords. Finally, many authentication systems are interactive by design, requiring explicit user input (e.g. QA prompts \cite{dandapat2015activpass} or pop-ups). Such overt mechanisms are unsuitable for the stealth and safety requirements of IPI detection.

\noindent
\textbf{Human activity recognition (HAR) and digital behavior study}: A fundamental distinction between HAR and detecting IPI-action lies in the objectives. Most HAR research focuses on \textit{physical movements} \cite{zhang2022deep, yang2025unsupervised} with sensory data such as inertial measurement units (IMUs). However, physical movements alone offer limited insight into \textit{what} the user is doing within an app. In contrast, detecting IPI actions~requires much higher-level semantic and contextual understanding that cannot be captured solely by motion  or vision-based sensing. This points to a different line of research---digital behavior study.

Recent studies in contextual and semantic digital behavior understanding on mobile have used app metadata and user-interface-trait data to analyze users' in-app activities. For example, \cite{songyan2024atool} proposed a tool that captures texts displayed on phone screens to analyze user psychological states. On top of screen content, other types of data have been explored for some general behavior understanding such as location, WiFi SSID and app usage \cite{xu2024autolife, kang2020automated, OTEBOLAKU201633}, providing diverse approaches for understanding human behavior on the smartphones. 

However, these two approaches fall short in the IPI context for two key reasons. \textit{First}, they mostly focus on general-purpose behavior summarization or contextual sensing and have no safety considerations to adapt to risky IPI scenarios, and do not incorporate safety-critical considerations required in IPI scenarios. \textit{Second}, many of these methods rely on data sources that seriously raise privacy concerns, such as screen display records, which are unsafe and impractical to collect in sensitive environments. These constraints make behavior understanding in the IPI context uniquely challenging.

To the best of our knowledge, no prior work has attempted to \textit{detect fine-grained IPI-related actions that users perform while interacting with smartphones, especially under stealth, safety, and privacy constraints}.

\noindent
\textbf{Need for IPI-tailored digital solutions.} The limitations discussed in this section motivate the need for a novel digital solution that 1) enables \textit{fine-grained and automated} sensing of IPI events, and 2) \textit{preserves user privacy} while 3) remaining \textit{subtle to minimize discovery and escalation}.

\section{PROBLEM FORMULATION}\label{sec-3-threatModel}

\subsection{Threat Model}
\label{subsection: threatmodel}

 We define an \underline{\textit{intimate partner}} broadly as someone, \add{regardless of gender}, with a close personal relationship with the device owner and shared or frequent access to the owner's smartphone. This includes current or former romantic partners, close friends, family members, and cohabiting individuals such as roommates.

We focus on smartphone-based IPI behaviors, and define the \deleted{\underline{\textit{victim}}}\add{\underline{\textit{owner}}} as the legitimate owner of the smartphone, and the \deleted{\underline{\textit{abuser}}}\add{\underline{\textit{intimate adversary}}} as the intimate partner who accesses and operates on the owner's device.

\add{\noindent\textbf{Target Scenario.} IPV has a complex landscape and varying levels of severity, ranging from moderate concerns to life-threatening physical violence~\cite{freed2018stalker, havron2019clinical}. Recognizing that no single technical tool can address this full spectrum, \REV{this work focuses on an early-stage IPI scenario in which a device owner is concerned about possible phone access by an intimate partner but lacks structured evidence about what has occurred on the device.} In such situations, a phone-side system such as \REV{\sys~can help collect records of candidate IPI-risk events by linking non-owner-like access with phone actions that are relevant to IPI risk. The non-owner may be the \textit{intimate adversary}, but \sys~does not infer relationship status, malicious intent, or whether an observed event constitutes IPV.} These records can support later review by trained support professionals, security clinics, or trusted organizations, helping them better understand the phone-side context and decide what follow-up steps may be appropriate for the owner's safety and situation.}

\noindent\textbf{Adversary's capabilities.} We make the following assumptions on  \MINOR{the} adversary's capabilities:

\begin{table}[t!]
    \small
    \centering
    \caption{Comparison between IPI and traditional cybersecurity attackers.}
    \begin{tabular}{@{}lll@{}}
        \toprule
        \textbf{Attacker Profile} & \textbf{Traditional} & \textbf{IPI} \\ 
        \midrule
        Tech skill & Advanced~\cite{sommer2010outside, aledhari2017protecting, fogla2006evading} & Less technical~\cite{tseng2020tools, roundy2020many} \\ 
        Physical Access & Mostly no~\cite{borky2018protecting} & Mostly yes~\cite{tseng2020tools} \\ 
        Passcode & Limited & Registered or guessable~\cite{tseng2020tools} \\
        Defense & Patches or updates & Security clinics~\cite{bellini2024abusive, tseng2022care, havron2019clinical} \\
        \bottomrule
    \end{tabular}
    \label{tab-ipvvstradition}
\end{table}

\noindent
\underline{1. Authenticated access}. As shown in Table \ref{tab-ipvvstradition}, unlike traditional attackers, intimate adversaries are assumed to have authenticated access to owners' devices, e.g., by registering their biometrics or having access to passwords. 

\noindent
\underline{2. User-interaction level operations}. Aligned with previous research \cite{freed2018stalker}, we assume that intimate adversaries have limited technical backgrounds, since intimate adversaries typically come from the general population compared to traditional cyber attackers. This means that abusive behaviors are restricted to UI-bound interactions, e.g., viewing content on the phone or modifying the configurations. 

\noindent
\underline{3. Adversary awareness of \sys}. Our baseline assumes that the intimate adversary is unaware of \sys on the phone, which is consistent with prior assumptions of the adversary's limited technical background. Nonetheless, we also analyze stronger and extreme yet rare cases (aware but passive; aware and actively probing) in Section~\ref{sec:stealthy_functioning} to demonstrate \sys's resilience to discovery attempts.

\noindent
\underline{4. Application scope}. \sys is designed to be \underline{app-agnostic}, unlike most prior adversary or intrusion detection studies that focus on finance \cite{bellini2023paying, fereidooni2023authentisense, cai2024famos}. Section \ref{sec-5-eval} evaluates six everyday apps across six representative categories (E-commerce, communication, social media, collaboration, music, and video streaming).

\noindent\textbf{Non-IPI behavior.} Prior threat models consider only harmful behaviors. However, a common scenario is that the current user is neither the owner nor the intimate adversary and is interacting with the device in a non-harmful way. This is common among friends, family members, or cultures where sharing personal devices is common (e.g., watching a video or performing an online search) \cite{karlson2009can, doerfler2024privacy}. Detecting this case as \emph{non-IPI} is essential for keeping AID’s false-positive rate low.

\subsection{\INTERVIEW{Formative Expert Interviews}}
\label{subsec:expert-interviews}

\INTERVIEW{Before developing the IPI behavior taxonomy and system design, we interviewed experts (n=5, Ph.D. students or Assistant Professors recruited from email lists and flyers) in technology-facilitated IPV, HCI, security, and privacy. Two had IPV-domain expertise through clinics or research publications (E1--E2), and three had HCI/security/privacy expertise in mobile systems, user studies, or safety-sensitive technologies (E3--E5). The interviews inform design principles for tech and clarify what kinds of early
technical evaluation are ethically and scientifically valid.}

\INTERVIEW{The interview asks experts how technology-facilitated IPV/IPI differs from traditional cybersecurity, what makes survivor support difficult, what device or account signals are useful in practice, and how a safer support tool should be designed and evaluated. The full question list can be found in Table~\ref{tab:interview-protocol} in Appendix~\ref{app:interview-questions}.}

\INTERVIEW{We analyzed the interview transcripts through a lightweight thematic analysis. We first extracted segments related to threat modeling, technical design, safety, privacy, participant recruitment, ground truth, and deployment. We then grouped related segments into recurring themes and used them to derive design principles and claim boundaries. Table~\ref{tab:expert-interview-themes} summarizes the main themes.}

\begin{table*}[t]
\centering
\footnotesize
\setlength{\tabcolsep}{4pt}
\caption{\INTERVIEW{Themes from expert interviews and their implications for IPI technology design. }}
\label{tab:expert-interview-themes}
\begin{tabular}{@{}p{0.18\textwidth}p{0.5\textwidth}p{0.3\textwidth}@{}}
\toprule
\textbf{\INTERVIEW{Theme}} & \textbf{\INTERVIEW{Representative expert quote}} & \textbf{\INTERVIEW{Design implications}} \\
\midrule

\INTERVIEW{IPI differs from conventional cybersecurity.}
&
\INTERVIEW{E1: ``Abusers... are often... physically co-located with the target... or have privileged access to their devices or accounts.''} 
\INTERVIEW{E3: ``The attacker is someone that... is close contact with you and know[s] a lot of information about you.''}
&
\INTERVIEW{Model IPI as close-person, phone-side access rather than only remote compromise; see Sections~\ref{subsection: threatmodel} and~\ref{subsec:formulating}.} \\

\midrule
\INTERVIEW{IPI can use ordinary end-user features.}
&
\INTERVIEW{E1: ``The abusers in these cases often use legitimate end-user features... the context under which they use them causes harm.''}
&
\INTERVIEW{Include ordinary UI-level phone actions in the taxonomy and behavior labels; see Sections~\ref{subsec:ipi_taxonomy} and~\ref{scalable}.} \\

\midrule
\INTERVIEW{Attribution, intent, and harm are hard to infer.}
&
\INTERVIEW{E2: ``It is hard to distinguish... the perpetrator and... the victim, because they could be acting from the same devices... [or] the same digital account.''} 
\INTERVIEW{E4: ``It is sort of hard to... define when it violates the other person's privacy.''}
&
\INTERVIEW{Treat outputs as candidate records, not proof of abuse, intent, relationship status, or harm; see Sections~\ref{subsec:formulating},~\ref{subsec:flagging}, and~\ref{subsec:limitation}.} \\

\midrule
\INTERVIEW{Automatic response can create safety risks.}
&
\INTERVIEW{E1: ``Whether you should revoke... an adversary session... is a lot more complex... [because of] safety planning.''} 
\INTERVIEW{E2: ``We cannot completely lock out the user... because that could be their only way of getting support.''}
&
\INTERVIEW{Avoid automatic blocking, forced account changes, visible alerts, or abuse claims; see Sections~\ref{subsec:flagging}, and~\ref{sec:stealthy_functioning}.} \\

\midrule
\INTERVIEW{Real IPV ground truth is very difficult to obtain and ethically sensitive.}
&
\INTERVIEW{E2: ``[Getting ground truth] in this situation is extremely hard...''} 
\INTERVIEW{E5: ``You cannot just recruit abusers. That makes it so hard to study the real abuser and survivor pairs.''}
&
\INTERVIEW{Frame controlled tasks as signal-feasibility tests, not evidence of real IPV dynamics; see Sections~\ref{subsec:dataset_settings} and~\ref{subsec:limitation}.} \\

\midrule
\INTERVIEW{Simulation or in-lab testing is a workaround.}
&
\INTERVIEW{E1: ``[Scenario-based testing] is quite removed from a real abuser, but it helps provide reasonable confidence on whether your system is doing, say, detection is working as intended.''}
\INTERVIEW{E2: ``... we should still do research in the direction of detecting abuse, for example, with anomaly detection, but exercise caution in claiming its utility for survivors when they are not part of the study.''}
\INTERVIEW{E4: ``For in-lab ones, you have control over the environment... [and] you can basically test edge cases.''}
&
\INTERVIEW{Use healthy close-pair tasks as a best-effort feasibility test enabled by the identity--action split, not as proof of survivor utility; see Sections~\ref{subsec:formulating},~\ref{subsec:dataset_settings}, and~\ref{subsec:limitation}.} \\

\midrule
\INTERVIEW{Survivor-facing validation remains a limitation and future work.}
&
\INTERVIEW{E2: ``Anything that's being built for the survivors should be built with the survivors.''}
\INTERVIEW{E4: ``In-lab testing... does not fully simulate what is going on in the real world.''}
\INTERVIEW{E5: ``If you can find local... organizations who provide support to these survivors... everything will be better.''}
&
\INTERVIEW{Limitation: current evaluation lacks survivor and clinic deployment. Future work should test survivor-facing use with survivors, advocates, and clinics; see Sections~\ref{subsec:limitation}--\ref{subsec:future}.} \\

\bottomrule
\end{tabular}
\end{table*}

\noindent
\textbf{\INTERVIEW{Interview takeaways for design.}}
\INTERVIEW{The expert themes shaped a narrow role for \sys. Because IPI often involves close-person access and ordinary phone features (E1, E3), \sys models phone-side IPI-risk evidence as a joint identity--action signal rather than as remote intrusion or malware alone (Section~\ref{subsec:formulating}). Because attribution, consent, intent, and harm require context outside sensor traces (E2, E4), \sys does not claim to determine IPI or decide interventions. Its output is a ranked candidate record (who, likely use, what action category, and when) for later interpretation by trained professionals. The safety implications of these themes (why \sys avoids automatic blocking, real-time alerts, and owner-facing outputs) are implemented in Sections~\ref{subsec:flagging} and~\ref{sec:stealthy_functioning}. The methodological implications on ``why we evaluate with healthy close pairs rather than survivors'' are discussed in Section~\ref{subsec:dataset_settings}.}

\subsection{Constructing IPI Behavior Taxonomy}
\label{subsec:ipi_taxonomy}

\begin{table}[h]
    \footnotesize
    \centering
    \caption{Taxonomy of Intimate Partner Infiltration (IPI) behaviors, classified at three granularities. The goal of \sys is to distinguish between categories (5-class), actions (9-class), and subactions (28-class) of IPI-related behaviors.}
    \label{tab:actionsubaction}
    \begin{tabular}{@{}lll@{}}
    \toprule
    \textbf{Category}      & \textbf{Action}               & \textbf{Subaction}                   \\ \midrule
    
    \multirow{4}{*}{Impersonation\cite{ceccio2023sneaky,bellini2023digital}} 
                           & \multirow{4}{*}{Send content} &    Send reviews \cite{freed2019my}                         \\
                           &                              & Send messages  \cite{10.1145/3637316,daffalla2023account,bellini2023digital}                       \\
                           &                              & Send emails                        \\
                           &                              & Comment   \cite{freed2018stalker}                            \\ \midrule
    \multirow{12}{*}{Leakage} 
                           & \multirow{5}{*}{View account} & View account settings \cite{10.1145/3637316,ceccio2023sneaky}                \\
                           &                              & Subscription details                  \\
                           &                              & Inspect order history                 \\
                           &                              & View browsing history \cite{havron2019clinical,freed2019my}            \\
                           &                              & View payment settings \cite{bellini2023digital}                \\ \cmidrule{2-3}
                           & \multirow{5}{*}{View content\cite{tseng2020tools}}  & View emails \cite{ramjit2024navigating}                          \\
                           &                              & See one's post history \cite{thomas2021sok}               \\
                           &                              & Watch history  \cite{havron2019clinical}                     \\
                           &                              & View messages \cite{daffalla2023account,ceccio2023sneaky,tseng2020tools}                        \\
                           &                              & Inspect files \cite{tseng2020tools,thomas2021sok}                        \\ \cmidrule{2-3}
                           & \multirow{2}{*}{Upload content}               & Upload photo \cite{ramjit2024navigating}                         \\
                           &                              & Upload video \cite{freed2018stalker}                         \\ \midrule
    \multirow{10}{*}{Modification} 
                           & \multirow{5}{*}{Alter account settings\cite{10.1145/3637316}}  & Change profile photo \cite{havron2019clinical}           \\
                           &                              & Change email \cite{ceccio2023sneaky,havron2019clinical}                         \\
                           &                              & Change username  \cite{thomas2021sok}                     \\
                           &                              & Change password \cite{ramjit2024navigating,daffalla2023account,bellini2023digital,tseng2020tools}                       \\
                           &                              & Change address  \cite{havron2019clinical}                         \\ \cmidrule{2-3}
                           & \multirow{2}{*}{Modify content\cite{stephenson2023abuse}} & Delete emails  \cite{thomas2021sok,freed2018stalker}                   \\
                           &                              & Modify music list   \cite{ramjit2024navigating}                  \\ \cmidrule{2-3}
                           & \multirow{3}{*}{Alter files\cite{ramjit2024navigating}}  & Add a file                            \\
                           &                              & Delete a file  \cite{freed2018stalker}                       \\
                           &                              & Modify a file                         \\ \midrule
    Software installation                & Software installation        & Software installation \cite{10.1145/3637316, tseng2020tools, zhang2025abusability}                \\ \midrule
    \multirow{1}{*}{NIO (Non-IPI-Other)} & NIO                      & NIO                               \\ 
    \bottomrule
    \end{tabular}
\end{table}

\noindent\textbf{Existing IPV taxonomy.} An expanding body of research underscores how intimate adversaries exploit technology to harm their intimate partners~\cite{tseng2020tools,tseng2022care,chatterjee2018spyware,freed2019my,freed2018stalker,freed2017digital,matthews2017stories}. The consequences are far-reaching, including financial harm~\cite{bellini2023paying}, escalating to physical confrontations, and even homicide~\cite{southworth2005high}.

Among the tactics explored by~\cite{freed2018stalker,tseng2020tools,woodlock2017abuse,bellini2023digital,ceccio2023sneaky}, three distinct types of technology-facilitated IPV have been identified: remote, proximate, and physical access. Remote tactics include actions such as distributed denial-of-service (DDoS) attacks, SMS bombing, unauthorized remote logins, and location tracking~\cite{freed2018stalker,stephenson2023s}. Proximate tactics involve methods requiring closer physical presence, such as the use of spy cameras, router monitoring, or other forms of nearby surveillance~\cite{ceccio2023sneaky}. Physical access tactics, on the other hand, encompass direct interactions with devices, such as installing spyware~\cite{chatterjee2018spyware} or creepware~\cite{roundy2020many}, accessing phone records, or deploying keylogging tools to monitor keystrokes. These tactics rely on distinct attack vectors.

\noindent\textbf{Methods and rationales for constructing IPI-tailored taxonomy.} In this work, we introduce a taxonomy specifically focused on the \textit{physical access} category of IPV, which we term \textbf{Intimate Partner Infiltration (IPI)}, shown in Table \ref{tab:actionsubaction}. 

To construct the taxonomy, we extracted 367 technology-facilitated IPV incidents from 15 top-tier HCI and security papers (2017--2025). Two authors jointly screened each incident against our developed threat model and kept 173 cases, which form the empirical backbone of our taxonomy. 

We then annotate every case along three axes: (i) intimate adversary's purpose -- \textit{device configuration, private content, public reputation, etc.}; (ii) mode of access -- \textit{viewing, modifying, sending, etc.}; and (iii) reported or observable viable traits on smartphones. Cross-combining yielded 28 subaction-traits, clustered into nine actions and five categories. All sub-actions/action/category behaviors are supported by at least one case
(2–6 typical); this hierarchy serves as both the taxonomy for IPI-related attacks and class labels for detecting \ipvact.

In Table \ref{tab:actionsubaction}, \add{we classify IPI behaviors at three levels of granularity: five high-level categories, nine action-level actions, and 28 fine-grained subactions. This hierarchy, derived from prior IPV literature, forms the basis for our behavior detection model.}\deleted{\textit{Impersonation} involves intimate adversaries pretending to be the owner by sending emails, messages, reviews, or comments to manipulate perceptions or harm reputations~\cite{tseng2020tools,woodlock2017abuse}. \textit{Leakage} refers to accessing or extracting private information, such as account settings, browsing history, messages, or files, and sometimes uploading content like photos or videos to violate privacy~\cite{ceccio2023sneaky}. \textit{Modification} includes altering device settings, changing account credentials, or modifying files, disrupting the owner’s sense of control and potentially causing emotional or psychological harm~\cite{bellini2023digital}. Lastly, \textit{Software installation} involves the installation of software on owner's device without consent, potentially enabling covert monitoring or control~\cite{woodlock2017abuse}.} Additionally, the \textit{NIO (Non-IPI-Other)} category includes general, risk-neutral behaviors that do not map to our IPI taxonomy and resemble normal device usage, such as watching videos.

\begin{table}[t!]
    \small
    \centering
    \caption{\MINOR{OS-level and physical signals analyzed by \sys and the required APIs}}\label{tab:modalitycontent}
    \begin{tabular}{@{}llll@{}}
    \toprule
    \textbf{Modality}                      & \textbf{Data type}                      & \textbf{Description}                     & \textbf{API used}\\ \midrule
    \multirow{8}{*}{IMU}                   & \multirow{5}{*}{Motion Data}           & Gyroscope                            & \multirow{8}{*}{SensorEvent}    \\
                                           &                                         & Accelerometer                            \\
                                           &                                         & Linear accelerometer                     \\
                                           &                                         & Magnetometer                             \\
                                           &                                         & Rotation vector sensor                   \\ \cmidrule{2-3}
                                           & \multirow{3}{*}{Environment Data}      & Proximity sensor                         \\
                                           &                                         & Pressure sensor                          \\
                                           &                                         & Light sensor                             \\ \midrule
    \multirow{7}{*}{Systems (SYS)}               & \multirow{2}{*}{Network Traffic}       & Upstream Bandwidth             & \multirow{2}{*}{TrafficStats}          \\
                                           &                                         & Downstream Bandwidth                     \\ \cmidrule{2-3}
                                           & \multirow{3}{*}{Energy Consumption}    & Current                  & \multirow{3}{*}{\shortstack[l]{BatteryManager\\FileReader}}                \\
                                           &                                         & Voltage                                  \\
                                           &                                         & Temperature                              \\ \cmidrule{2-3}
                                           & \multirow{2}{*}{Memory Utilization}    & Memory used                & \multirow{2}{*}{\shortstack[l]{ActivityManager\\UsageStatsManager}}              \\
                                           &                                         & \#App usages in memory                   \\ \midrule
     \multirow{2}{*}{Interaction (INT)}                            & \multirow{2}{*}{Screen Interaction}    & Interaction rate           & \multirow{2}{*}{AccessibilityEvent}              \\
                                           &                                         & Interaction event                        \\ \midrule
    Application (APP)                            & Application Activeness                 & Foreground app name          & UsageStatsManager            \\ \bottomrule
    \end{tabular}
\end{table}

\begin{table}[t]
  \centering
  \footnotesize
  \caption{Conceptual mapping between IPI-related behavior detection
           and modality-level OS signals. 
           \cmark = strongly indicative; 
           \textcolor{orange}{$\sim$} = plausibly indicative; 
           \xmark = minimal / none.}
  \label{tab:action_modality_map}
  \begin{tabular}{@{}lcccc@{}}
    \toprule
    \textbf{Target} & \textbf{IMU (Motion)} & \textbf{SYS (Systems)} & \textbf{INT (Interaction)} & \textbf{APP (Application)}\\
    \midrule
    Identify user                        & \cmark & \cmark & \textcolor{orange}{$\sim$} & \xmark \\
    \midrule
    Send content                         & \xmark & \textcolor{orange}{$\sim$} & \cmark & \cmark \\
    View account/content                 & \xmark & \xmark & \cmark & \cmark \\
    Upload content                       & \xmark & \textcolor{orange}{$\sim$} & \cmark & \cmark \\
    Alter account settings               & \xmark & \cmark & \cmark & \cmark \\
    Modify content                       & \xmark & \textcolor{orange}{$\sim$} & \cmark & \cmark \\
    Alter files                          & \xmark & \textcolor{orange}{$\sim$} & \cmark & \cmark \\
    Software installation                & \xmark & \textcolor{orange}{$\sim$} & \cmark & \cmark \\
    NIO (benign use)                     & \xmark & \xmark & \xmark & \xmark \\
    \bottomrule
  \end{tabular}
\end{table}

\subsection{Mapping IPI-related Behaviors to OS signals.} 
\label{subsec-map-to-os-signal}

\MINOR{\add{To systematically bridge the selected behaviors with concrete system design, we explored incorporating four high-level signal modalities, as summarized in Table~\ref{tab:modalitycontent}. We developed a heuristic mapping between these modality-level OS signals and IPI-related behavior, which we summarize in Table~\ref{tab:action_modality_map}. All four modalities are retrievable via standard Android APIs and system services without requiring privileged access, ensuring practical deployability of \sys (Section~\ref{subsec:os_compatibility})}.}

\noindent
\MINOR{\textbf{Uncovering complementary modalities through ablations.} To study the efficacy of each modality on the IPI-risk detection pipeline, we conducted ablation studies where we vary the data streams used as input for identifying users and behaviors. Tables~\ref{tab:mod1modality} and~\ref{tab:mod2modality} show these results. We see that incorporating all streams of data does not yield the best performance, which contradicts intuition, and that carefully selecting the most relevant modalities performs better. IMU and SYS signals, which capture a user's physical micro-habits, proved to be the most predictive indicators of user identity. In contrast, INT and APP signals, reflecting on-screen actions, were the most effective for identifying behaviors. These results support our developed mapping, summarized in Table~\ref{tab:action_modality_map}. This clear separation of concerns serves as the foundational blueprint for our system's architecture with its two key components: user-identity and behavioral action, which we formally define next.}

\begin{table*}[]
\centering
\caption{\MINOR{F1 scores (mean ± std) for user identification across different modality combinations. Best score in each row is \textbf{bolded}.}}
\label{tab:mod1modality}
\begin{adjustbox}{width=\linewidth}
\begin{tabular}{lcccccccc}
\toprule
\textbf{Pair Type} & IMU & SYS & INT & APP & IMU+APP & IMU+SYS+APP & ALL & \textbf{IMU+SYS (final)} \\
\midrule
Partner & \smallpm{0.982}{0.083} & \smallpm{0.637}{0.022} & \smallpm{0.511}{0.072} & \smallpm{0.621}{0.222} & \smallpm{0.978}{0.113} & \smallpm{0.893}{0.083} & \smallpm{0.928}{0.317} & \textbf{\smallpm{0.998}{0.002}} \\
Stranger & \smallpm{0.935}{0.121} & \smallpm{0.559}{0.228} & \smallpm{0.554}{0.176} & \smallpm{0.651}{0.146} & \textbf{\smallpm{0.999}{0.001}} & \smallpm{0.844}{0.092} & \smallpm{0.936}{0.063} & \smallpm{0.977}{0.041} \\
Overall & \smallpm{0.959}{0.090} & \smallpm{0.598}{0.174} & \smallpm{0.532}{0.202} & \smallpm{0.636}{0.247} & \smallpm{0.985}{0.022} & \smallpm{0.853}{0.178} & \smallpm{0.932}{0.074} & \textbf{\smallpm{0.987}{0.031}} \\
\bottomrule
\end{tabular}
\end{adjustbox}
\end{table*}

\begin{table*}[htbp!]
\small
\centering\caption{\MINOR{Top-k accuracies (mean ± std) for \ipvact classification across different modality combinations. Best score in each row is \textbf{bolded}.}}
            \label{tab:mod2modality}
\begin{tabular}{llcccccc}
\toprule
\textbf{Classification Level} & \textbf{Accuracy} & \textbf{INT} & \textbf{APP} & \textbf{IMU} & \textbf{SYS} & \textbf{ALL} & \textbf{INT+APP (final)} \\
\midrule
\multirow{3}{*}{\textbf{Category (5 classes)}}
& Top-3 & \smallpm{0.945}{0.007} & \smallpm{0.917}{0.025} & \smallpm{0.832}{0.035} & \smallpm{0.838}{0.027} & \smallpm{0.907}{0.035} & \textbf{\smallpm{0.959}{0.010}} \\
& Top-2 & \smallpm{0.850}{0.015} & \smallpm{0.754}{0.047} & \smallpm{0.619}{0.058} & \smallpm{0.649}{0.045} & \smallpm{0.772}{0.057} & \textbf{\smallpm{0.885}{0.025}} \\
& Top-1 & \smallpm{0.613}{0.015} & \smallpm{0.518}{0.056} & \smallpm{0.367}{0.074} & \smallpm{0.370}{0.027} & \smallpm{0.534}{0.057} & \textbf{\smallpm{0.652}{0.039}} \\
\midrule
\multirow{3}{*}{\textbf{Action (9 classes)}}
& Top-3 & \smallpm{0.848}{0.017} & \smallpm{0.795}{0.049} & \smallpm{0.526}{0.067} & \smallpm{0.600}{0.052} & \smallpm{0.775}{0.066} & \textbf{\smallpm{0.900}{0.035}} \\
& Top-2 & \smallpm{0.748}{0.022} & \smallpm{0.639}{0.054} & \smallpm{0.388}{0.074} & \smallpm{0.452}{0.055} & \smallpm{0.637}{0.070} & \textbf{\smallpm{0.808}{0.045}} \\
& Top-1 & \smallpm{0.553}{0.023} & \smallpm{0.446}{0.053} & \smallpm{0.240}{0.071} & \smallpm{0.278}{0.045} & \smallpm{0.446}{0.073} & \textbf{\smallpm{0.596}{0.054}} \\
\midrule
\multirow{3}{*}{\textbf{Subaction (28 classes)}}
& Top-3 & \smallpm{0.718}{0.028} & \smallpm{0.601}{0.066} & \smallpm{0.264}{0.066} & \smallpm{0.307}{0.064} & \smallpm{0.591}{0.109} & \textbf{\smallpm{0.795}{0.046}} \\
& Top-2 & \smallpm{0.620}{0.032} & \smallpm{0.481}{0.048} & \smallpm{0.189}{0.045} & \smallpm{0.236}{0.057} & \smallpm{0.488}{0.103} & \textbf{\smallpm{0.709}{0.050}} \\
& Top-1 & \smallpm{0.462}{0.040} & \smallpm{0.350}{0.043} & \smallpm{0.109}{0.033} & \smallpm{0.148}{0.039} & \smallpm{0.345}{0.088} & \textbf{\smallpm{0.566}{0.058}} \\
\bottomrule
\end{tabular}
\end{table*}

\subsection{Formulating IPI-risk Event Detection}
\label{subsec:formulating}

\noindent
\textbf{Problem statement.} The primary goal is: given a time-series multi-modal window
\(x\in\mathbb{R}^{T\times d}\) (T time steps, d features),
learn a mapping \(f:x\!\rightarrow\!y\) where
\(y\in\{0,1\}\) indicates Intimate Partner Infiltration (IPI) risk.

\noindent
\textbf{Joint formulation.} Standard anomaly-detection or continuous-authentication \cite{10.1145/3130975,10.1145/3351243} produces a single outlier score, implicitly conflating \emph{who} is holding the phone with \emph{what} they are doing; this yields false alarms under benign sharing. Instead, we reformulate the detection as a structured problem considering two factors jointly:

\begin{enumerate}[leftmargin=*]
    \item \textbf{User identity}: $u \in {0,1}$ denotes the identity of the current phone user, where $u=0$ represents the \textit{victim}, i.e., the legitimate device owner in our target scenario, and $u=1$ represents a \textit{non-owner} who may access the device without the owner’s explicit authorization.

    \item \textbf{Behavior}: $b \in \{0, 1\}$ denotes whether the observed phone-side behavior is IPI-risk related according to our predefined taxonomy (Table~\ref{tab:actionsubaction}), where $b = 0$ represents IPI-neutral or risk-neutral behavior, and $b = 1$ represents an IPI-risk interaction that warrants flagging for later review.
\end{enumerate}

\begin{equation}
y=\mathbb{I}\bigl[u=1 \;\land\; b=1\bigr]
\label{eqn:ipi_formulation}
\end{equation}

As shown in Equation~\ref{eqn:ipi_formulation}, an event ($y$) is IPI-positive only when
\emph{both} the actor is not the legitimate owner and the behavior is IPI-related.
The detection task therefore reduces to jointly estimating
\(u\) and \(b\) from \(x\); Section \ref{sec-4-sysdesign} describes our dual-branch architecture
that realizes this joint reasoning, while Section~\ref{sec-5-eval} demonstrates the advantages of this paradigm compared to other state-of-the-art approaches.

\section{\sys~DESIGN}\label{sec-4-sysdesign}

\begin{figure}[t!]
    \centering\includegraphics[width=\columnwidth]{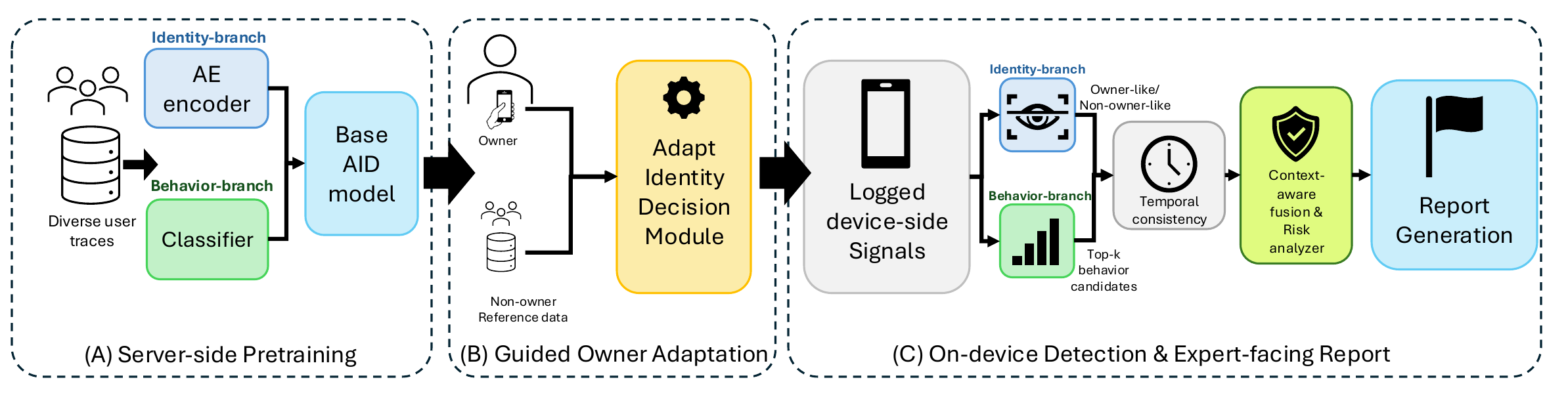}%
\caption{\MINOR{\sys workflow and system architecture. 
(A) Server-side pretraining learns an identity encoder and behavior classifier from diverse user traces. 
(B) During guided owner adaptation, \sys uses a short owner calibration session and non-owner reference data to adapt the identity decision module, while keeping the behavior branch fixed. 
(C) During deployment, logged device-side signals are processed by the identity and behavior branches; temporal consistency and context-aware fusion combine owner-likeness evidence and ranked behavior candidates to support expert-facing report generation. 
The report is intended to flag candidate IPI-risk events for later expert review only.}}
\label{fig:aid_workflow}\label{fig:workflow}
\end{figure}

\subsection{Workflow Overview}

Figure~\ref{fig:workflow} shows \sys's workflow. It takes in fixed-length time-series sequences, where signals from multiple modalities are concatenated and min-max normalized per feature.

\noindent
\textbf{Pretraining.} \sys is designed in a dual-branch manner based on our formulation in Section~\ref{subsec:formulating}: 1) \emph{Identity}-branch utilizes an autoencoder (AE) to confirm if the user is the owner and 2) the \emph{Behavior}-branch classifies behaviors into our taxonomy defined in Section~\ref{subsec:ipi_taxonomy}. Finally, 3) \sys fuses the output of these two branches through a context-aware risk analyzer that leverages $T$, a learned thresholding parameter. Before deploying \sys, we pretrain its dual-branch architecture on data collected from a diverse user group.

\noindent
\textbf{Adaptation.} When clinics decide to deploy \sys on an owner's phone, \sys fine-tunes the decision network in the identity-branch, with five minutes of the owner's usage data collected during an initial calibration phase to enable \sys to distinguish phone usage by the owner and non-owners.

\noindent
\textbf{Detection and reporting.} This stage occurs on-device post-adaptation. It uses the adapted system to identify anomalous users and suspicious behaviors indicative of IPI events. The suspicious results, which contain time stamps, app usage, and contextual evidence, are silently logged and securely reported to authorized support channels (i.e., clinic experts, see Section~\ref{subsec:deploymentwithclinics} for details).

\subsection{Pretraining Stage}\label{scalable}

\begin{figure}[h!]
    \centering
    \begin{subfigure}[b]{0.7\linewidth}
        \centering
        \includegraphics[width=1\columnwidth]{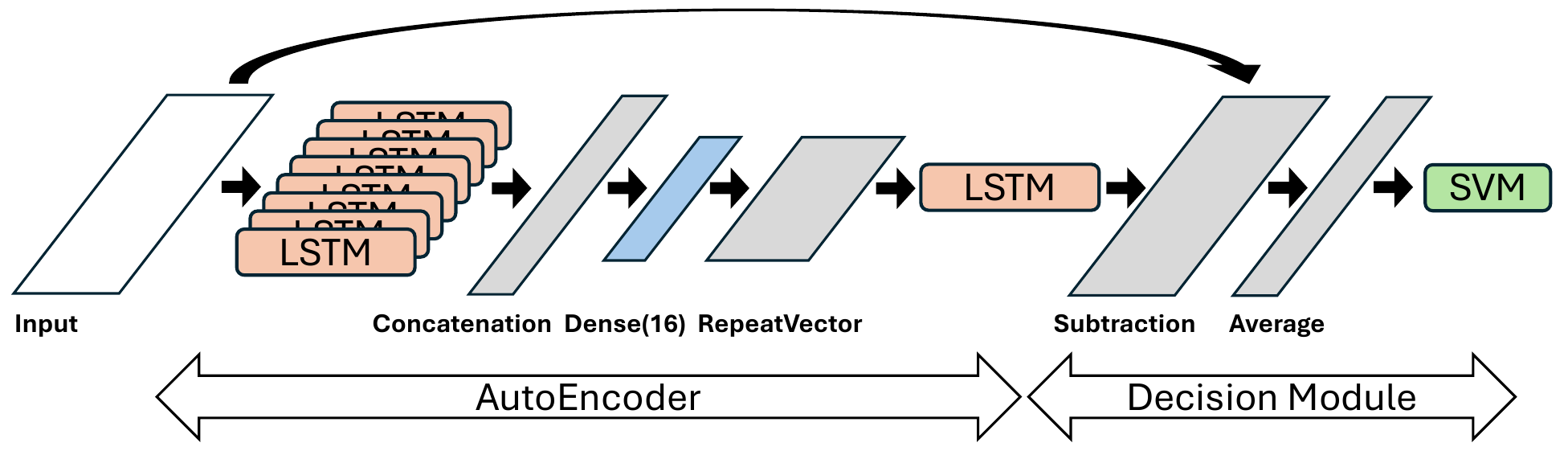}
        \caption{User Identity Branch Architecture.}
        \label{fig:Mod1}
    \end{subfigure}
    \hfill
    \begin{subfigure}[b]{0.7\linewidth}
        \centering
        \includegraphics[width=\columnwidth]{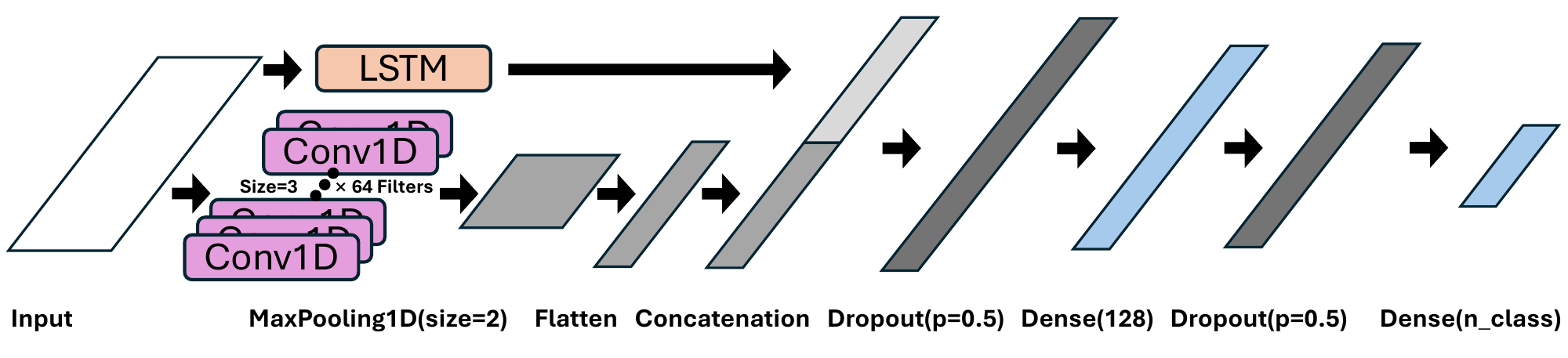}
        \caption{Behavior Branch Architecture.}
        \label{fig:Mod2}
    \end{subfigure}
    \caption{\sys's model architectures for a) identifying the phone's owner and b) detecting \ipvact.}
    \label{fig:ModelArchitecture}
\end{figure}

\noindent
\textbf{User identity-branch.} There is large variability between different users, so it is difficult to train a single model that can distinguish one user from another. As such, we adopt a fine-tuning scheme that adapts to the phone's owner. This branch relies on IMU and SYS modalities, which capture users’ micro-habits and system-level patterns that are most indicative of user identity (as discussed in Section \ref{subsec-map-to-os-signal}).

As shown in Figure~\ref{fig:Mod1}, we propose a multi-head LSTM-based Auto-Encoder that extracts latent features which are analyzed by an SVM-based decision module to determine the identity of the user. The SVM decision module is adapted to new users and phones with a small amount of labeled data, which is discussed in Section~\ref{subsection:adaptprocess}.

On the other hand, the Auto-Encoder is used to learn general features in behaviors that are common across the vast majority of users and is trained with unlabeled data in an unsupervised manner. The encoder consists of $H$ parallel LSTM heads, each receiving the same input sequence and encoding it into a distinct latent representation. These embeddings are then concatenated into a single feature vector, which is then passed to a single LSTM decoder to reconstruct the original input sequence. This multi-head structure encourages the encoder to capture diverse signals in parallel.

\noindent
\textbf{Behavior-branch.} While there is significant work in the area of human activity recognition (HAR) on smartphones, detecting \ipvact is significantly different. Rather than viewing phones and wearables as a system that a person interacts with, HAR leverages these devices as a general-purpose sensor to detect motions and actions (e.g., step counting, motion tracking) \textit{while a person is not directly interfacing with the smartphone}. This often involves heavy use of physical sensors \MINOR{on the} device (e.g., camera and IMU). In contrast, determining \ipvact involves \textit{detecting actions a user performs while interfacing directly with the phone} (e.g., changing passwords or browsing private information). As such, the modalities and techniques used need to be adjusted. This branch relies on INT and APP modalities, which capture user interactions and foreground application contexts that best reflect behaviors (see Section \ref{subsec-map-to-os-signal}).

Given the challenges of obtaining labeled data post-installation, we implement a server-side training architecture and deploy only the trained detection model on-device for direct inference without further user-side training. Recent advances in HAR have explored various architectures, from CNNs~\cite{wan2020deep} and LSTMs~\cite{mekruksavanich2021deep} to Transformers~\cite{ek2022lightweight, mohapatra2026maestro}. Inspired by the success of hybrid models in HAR~\cite{zhang2022deep}, we propose an LSTM-CNN architecture for classifying \ipvact. As shown in Figure~\ref{fig:Mod2}, our model processes input sequences through parallel paths: an LSTM layer with 64 units and a one-dimensional convolutional layer (kernel size=3, 64 filters). The CNN output undergoes max pooling and flattening before concatenation with the LSTM output. These features pass through two dropout-dense blocks for behavior classification. To train the classifier, we use a categorical cross-entropy loss:

\begin{equation}
\mathcal{L}_{\text{cls}} = -\sum_{i=1}^{C} b_i \log(\hat{b}_i)
\end{equation}

\noindent
where \( b_i \in [0, 1] \) is a soft label distribution over \( C \) classes with \( \sum_i b_i = 1 \), and \( \hat{b}_i \) is the predicted class probability.

\noindent
\textbf{Context-aware fusion and risk estimator.} To decide whether a window should be flagged as a candidate IPI-risk event, we design a context-aware risk analyzer that calculates a decision score $S$ based on the outputs from both the identity and behavior branches. This score is based on the confidence and margin between NIO and suspicious predictions. We first check the user identity score. If the user is verified (i.e., $\hat{u} = 0$), the event is labeled as safe regardless of behavior. Otherwise, we apply the fusion rule over the behavior predictions. Specifically, we define:

\begin{equation}
S = \left( \frac{\hat{b}_{\text{NIO}}}{\sum_{j \ne \text{NIO}} \hat{b}_j + \epsilon} \right) \cdot \left( \hat{b}_{\text{NIO}} - \frac{1}{k-1} \sum_{j \in \mathcal{T}_k \setminus \{\text{NIO}\}} \hat{b}_j \right)
\end{equation}

\noindent
where \( \hat{b}_j \) denotes the predicted probability for class \( j \), \( \mathcal{T}_k \) is the set of top-\(k\) predicted classes, \( \text{NIO} \) is the index of the NIO class, and \( \epsilon \) is a small constant to prevent division by zero.

We then determine whether an IPI event has occurred based on a threshold \( T \) over the score \( S \):
\begin{equation}
\hat{y} = 
\begin{cases}
0 & \text{if } S > T \\
1 & \text{otherwise}
\end{cases}
\end{equation}

\noindent
To determine the optimal threshold \( T \) for IPI decision-making, we perform a grid search on a validation set and select the value that maximizes the F1 score of IPI event detection. This approach is more interpretable and achieves higher performance than fusing both branches in a black-box manner, as we will show in Section~\ref{sec-5-eval}.

\subsection{New User Adaptation}
\label{subsection:adaptprocess}

\noindent\textbf{Few-shot learning for efficient adaptation.} Model training is typically resource-intensive and time-consuming. In our dual-branch architecture, the  behavior-branch performs inference using a trained model, while the identity-branch requires fine-tuning for user adaptation. To minimize this overhead, we implement the few-shot learning scheme that enables efficient adaptation with minimal data. During pretraining, the Auto-Encoder is trained on a diverse dataset, capturing general patterns of mobile interaction. This knowledge enables the model to rapidly adapt to a new user's specific interaction patterns through fine-tuning. Our experiment (Appendix \ref{appendix:pretrain_finetune}) demonstrates that random sampling of just 20 samples during the calibration session is sufficient for fine-tuning, achieving high detection accuracy. This eliminates the need to process the full dataset, significantly reducing computational and time costs. The resulting system is both practical and efficient for real-world deployment.

\noindent
\textbf{Fine-tuning for user adaptation} When the owner installs the application for the first time, \sys guides the owner through a short calibration procedure of 5 minutes, where the owner uses the device naturally. This session provides sufficient data for fine-tuning the classifier to adapt to the owner's unique behavioral patterns. The fine-tuning dataset is constructed by selecting a subset of newly collected data from the owner and pre-stored data from other users (labeled as negative examples and anonymized to prevent any potential IPI leakage). To ensure the balance during classifier training, we maintain a 1:1 ratio between the selected samples from the owner and non-owner users.

After selecting data for adaptation, we use the frozen pretrained Auto-Encoder to extract \textit{difference vectors}. For each input sequence \( \mathbf{x}_{1:T} \), we compute its reconstruction \( \hat{\mathbf{x}}_{1:T} \), and derive a difference vector by averaging the difference over the temporal axis:

\begin{equation}
\mathbf{d} = \frac{1}{T} \sum_{t=1}^{T} \left( \mathbf{x}_t - \hat{\mathbf{x}}_t \right)
\end{equation}

\noindent
This results in a fixed-length vector \( \mathbf{d} \in \mathbb{R}^d \), which captures the discrepancy between the input and its reconstruction over time. A decision model---a one-class SVM---is trained using the difference vectors from the selected owner's and non-owner users' samples to establish a decision boundary. During inference, each difference vector is classified as either $\hat{u} = 0$ (legitimate device owner) or $\hat{u} = 1$ (intimate adversary/non-owner).

\subsection{\REV{Candidate Event Flagging and Report Generation}}
\label{subsec:flagging}
After adaptation is complete, \sys~begins logging and inference on logged data for suspicious IPI behaviors.

\noindent
\textbf{Temporal consistency module (TCM) for stable detection.}  To further enhance reliability, we introduce a \textit{Temporal Consistency Module (TCM)} as a post-processing stage applied to both user identification and behavior classification before context-aware fusion. The design of TCM is motivated by the observation that isolated mispredictions often occur amidst a sequence of correct and consistent outputs. Leveraging temporal continuity, TCM suppresses such outliers and ensures stable predictions over time.

\noindent
\textit{1. Identity-branch.} We extract temporal features (rolling mean and standard deviation) from prediction scores, which are input into a k-means clustering model (k = 2) that partitions the prediction space into regions corresponding to true users and potential intimate adversaries, serving as a temporal post-processing step to stabilize one-class SVM outputs. A temporal voting mechanism then aggregates predictions within a sliding window to reduce variability. This approach significantly reduces false positives while maintaining high detection sensitivity, which we will show in our experiments.

\noindent
\textit{2. Behavior-branch.} Post-processing for behavior classification adopts a simpler rolling window average to smooth fluctuations in predictions and reduce temporal variations. 

\noindent\textbf{Asynchronous detection.} While deep learning models are effective for accurate detection, running them continuously on mobile devices can lead to excessive resource consumption, overheating, and degraded user experience~\cite{sui2026modality}. Unlike traditional user authentication systems, which prioritize real-time processing and alerts---potentially notifying intimate adversaries---\sys performs detection asynchronously at night when the device is idle and charging. This design avoids real-time computational overhead, with the only daytime burden being lightweight logging of interaction signals, which is power-efficient (0.12\% battery drain over 60 minutes at 20~Hz sampling rate, 0.30\% at 40~Hz, 3.02\% at 130~Hz---well beyond typical operating frequency).


\definecolor{aidreportbg}{RGB}{226,238,252}
\definecolor{aidgreen}{RGB}{0,110,0}
\definecolor{aidred}{RGB}{180,0,0}


\begin{table*}[t]
\centering
\caption{\MINOR{Example expert-facing AID report. For non-owner-like windows, AID retains timestamp, app context, ranked top-\(k\) behavior evidence, and the final report flag for later expert review. For owner-like windows, app and behavior evidence are suppressed for privacy.}}
\label{tab:aid_report_example}

\resizebox{\textwidth}{!}{
\begin{tabular}{lllllll}
\toprule
Time & App & Identity evidence & Top-1 behavior & Top-2 behavior & Top-3 behavior & Report flag \\
\midrule
\makecell[l]{10/10/24\\15:38:39} &
Suppressed &
Owner-like &
Not retained &
Not retained &
Not retained &
\makecell[l]{Suppressed by\\privacy rule} \\

\makecell[l]{10/10/24\\18:32:47} &
YouTube &
Non-owner-like &
NIO (0.76) &
Upload content (0.11) &
View content (0.07) &
No risk flag \\

\makecell[l]{10/10/24\\19:48:27} &
Slack &
Non-owner-like &
Alter account settings (0.56) &
View content (0.32) &
NIO (0.05) &
\makecell[l]{Candidate\\IPI-risk event} \\
\bottomrule
\end{tabular}
}

\end{table*}

\noindent\textbf{Report generation.}
\MINOR{Detection results are compiled into a secure clinic-facing report, illustrated in Table~\ref{tab:aid_report_example}}. Each retained non-owner-like window includes the timestamp, app context, owner-likeness state, ranked top-\(k\) behavior candidates with scores, and the final report flag. Owner-like windows suppress app and behavior evidence to reduce privacy exposure. The report remains securely stored on the device until it is reviewed by trained support professionals, and its output is interpreted as device-side evidence for later review rather than as an automatic determination of abuse, intent, or relationship context.

\noindent
\textbf{Interpreting results with context.}
Unlike traditional human activity recognition (HAR), which leverages IMU or camera data to track physical movements, \sys operates under a more constrained and privacy-preserving setting. It infers behaviors from indirect signals—such as app usage and touch interaction—where distinct actions may appear similar (e.g., typing a password vs. searching YouTube). To accommodate such ambiguity, \sys logs the top-k behavior predictions for each anomalous user session, offering a broader context to support cautious and informed interpretation by security professionals.

\subsection{\COMPATIBLE{Implementation and OS Compatibility}}
\label{subsec:os_compatibility}

\MINOR{We implement \sys as a root-free Android app built in Android Studio, relying solely on standard SDK calls.}

\noindent
\MINOR{\textbf{Multi-modal data.} \sys samples and records four categories of modalities (Table~\ref{tab:modalitycontent}) to obtain a comprehensive view of how the user is interacting with the phone. These modalities are complementary in improving detection performance, as detailed in Section~\ref{subsec-map-to-os-signal}.}

\noindent
\MINOR{\textbf{APIs and permissions requirement.} All APIs are officially supported in the Android SDK and are compatible with a wide range of device models without requiring root access. The APIs used for each modality are summarized in Table \ref{tab:modalitycontent}. Additionally, we use the Accessibility Service API \cite{google2024accessibility} to register \sys~as a background service, enabling safe and stealthy monitoring (details in Section \ref{subsec:safety_design}). To function properly, \sys requires the following permissions: BIND\_ACCESSIBILITY\_SERVICE, READ/WRITE\_EXTERNAL\_STORAGE, and PACKAGE\_USAGE\_STATS. Upon installation, users are guided by a clinic helper to grant these permissions and launch the accessibility service. Then \sys~runs in the background  without appearing on the home screen or on the app-switcher page.}

\COMPATIBLE{To address compatibility with recent Android releases, we additionally tested AID on Android 16 and confirmed that its core logging and detection pipeline remains operational. However, the required Accessibility Service and usage-access permissions must be explicitly enabled during the clinic-guided setup process, rather than silently or automatically granted by the OS.}
\section{STRATEGIES FOR SAFETY AND STEALTHINESS}
\label{sec:stealthy_functioning}

\add{While \sys~incorporates several built-in stealth mechanisms that we detail in this section, we acknowledge that achieving perfect soundness and persistent undetectability against a motivated adversary remains a significant and largely orthogonal challenge in itself. We identified several viable security enhancements and discuss them in Section~\ref{risk_discussion}.}

\subsection{\add{Responsible} Deployment \add{and Usage Model}\deleted{of \sys with Security Clinics}}
\label{subsec:deploymentwithclinics}

\add{The deployment of \sys~is guided by a safety-first principle that acknowledges the risk in every phase of IPI conflict. To mitigate this, we propose a deployment model centered on a ``gatekeeper'' architecture, where the tool is dispatched and its outputs are interpreted with the guidance and control of trusted and professional entities instead of individual users.}

\add{The deployment workflow is as follows: An at-risk individual consults with a professional from a trusted organization on-site (e.g., a security clinic, a non-profit support group). After an initial assessment of the owner's current circumstances and with consensus between individual and professional, they facilitate the installation of \sys~on the individual's device. This process includes sideloading a signed APK. Then the professional guides the owner through the setup of the tool including granting the required permissions, launching through the Accessibility Service page, and completing a five-minute adaptation process for the user identity branch (Section \ref{subsection:adaptprocess}) where the owner uses the device in a natural way to let the device recognize the normal pattern.}

\add{Once installed and set up, \sys~operates as a secure ``black box'' on the device silently. Logs are saved to the app's private, inaccessible directory. Crucially, access to these logs is limited to the trusted professional. This is to ensure the evidence is only accessible when the individual is in a physically and emotionally safe environment. Then the professional can leverage these objective findings to collaboratively develop personalized, safer intervention strategies with the individual.}

\add{We believe this deployment model is a safer and intentionally conservative approach compared to making it publicly available. It mitigates the risk of disclosure, such as by force or under coercion by the adversary.}

\subsection{Safety Design}
\label{subsec:safety_design}

\noindent\textbf{Review of risky data sources.} \sys should be ``invisible'' to intimate adversaries while also being privacy sensitive towards the owner\deleted{of the phone}. As such, we discard data streams\deleted{only available} through APIs that violate these principles. We identified four such signals\deleted{, which are not present in \sys}:

\noindent
\underline{1. Touch} is available through the MotionEvent API and provides information about the coordinates of a user's touch, swipe direction, pressure etc. However, accessing this API in the background places a floating icon on the screen and triggers mandatory system alerts.

\noindent
\underline{2. Camera and Face ID.} \deleted{While cameras can accurately identify the current user, they} \add{They} record highly sensitive information such as the owner's face\deleted{which results in severe privacy breaches if leaked}. Also, activating the camera in the background triggers mandatory notifications, which alerts intimate adversaries. \deleted{Another potential solution is to directly leverage the camera or biometric sensors to perform user authentication, but}\add{For Face ID,} intimate adversaries often bypass it due to prior registration\deleted{or intimate knowledge of their partners}. Besides, it typically requires the user to look at the device with proper alignment to the sensors, which is impractical\deleted{for stealthy detection scenarios where} \add{as} the device may be used from various angles. \deleted{Furthermore, Biometrics and Face ID provide no defense, monitoring, or privacy against intimate adversaries who gain unfettered access to the phone, while \sys provides safe and stealthy monitoring of potential IPI instances after login.}

\noindent
\underline{3. Audio.} Similar to the camera, using audio through the MediaRecorder API raises significant privacy concerns and triggers mandatory on-screen notifications that can alert intimate adversaries.

\noindent
\underline{4. Location.} While location tracking through the LocationManager API\deleted{, GPS, and wireless signals} could provide useful information, it reveals sensitive details about the user’s \deleted{habits or frequent}locations. Additionally, as with camera use and audio recording, continuous location tracking can trigger system notifications and drain the battery, making it less suitable.

\noindent\textbf{User-interface deception.} We crafted our Android application to run in the background as a service, without a UI or application icon, and randomize the name. Moreover, we leverage the Accessibility Service~\cite{google2024accessibility}, a built-in function of Android OS that enables \sys to launch without clicking on the application icon and remain absent from the ``Recent Apps'' page to enhance stealthiness and prevent adversaries from terminating \sys there.

\noindent\textbf{Local inference and secure communication.} \sys performs inference and fine-tuning on-device, which eliminates the need to send logging data to a remote server, reduces the chance of data leaks, and enhances data privacy. Furthermore, we take advantage of the IPV context to further reduce \sys's energy footprint. Because we avoid alerting users to prevent discovery by intimate adversaries, \textit{\sys does not need to provide insights continuously and in real time}. As such, during the day, \sys collects and logs traces, and analyzes the collected data at night, when the device is likely idle and connected to power. Furthermore, considering that some intimate adversaries may have access to the home router and could monitor network traffic, \sys ensures that any transmitted messages are encrypted at the application layer, providing an additional layer of protection against network-level surveillance.

\subsection{What If the Intimate Adversary is Aware of \sys?}
\label{subsec:intimate adversary_is_aware}

Although the average IPV attacker may not have the same technical background as a cyber attacker, there is a chance that intimate adversaries may become aware of \sys, even with the safety measures employed. Here we discuss scenarios reflecting the degree of the intimate adversary's awareness and technical knowledge.

\noindent
\textbf{1. Intimate adversary is aware of \sys's existence}, but \MINOR{does not know} how to stop \sys. Here, \sys is still feasible, as the intimate adversary cannot prevent \sys. However, since \sys functions as a background service, a more informed intimate adversary can kill \sys through the services or Apps menu in Settings. While this can stop \sys, doing so reveals a clear IPI intention, which can be used to inform owners and security experts.

\noindent
\textbf{2. Intimate adversary is aware of \sys's capabilities and attempts to mimic the owner's behavior.} If an intimate adversary knows that \sys detects IPI-related behaviors, s/he may attempt to mimic the behavior of their partner to bypass detection. While this may work initially, many works have shown that it is difficult to consciously mimic the behaviors and biometrics of another person for extended periods~\cite{khan2020mimicry,xie2023fingerslid}. As such, over the long run, we anticipate that \sys would still capture unauthorized intrusions and IPI behavior. Moreover, once mimicry fails, the resulting deviations in sensor readings may further amplify anomalies, marking the session as suspicious and increasing the likelihood of recognition.

\done{\weisi{what do we do with the sharing with clinics if \sys~could be used without clinics?}}

\noindent
\textbf{3. Can \sys be used against the owner?} \deleted{Source code will not be made publicly available and data is only shared with clinics or other professional organizations and platforms.}\add{Access to the logs is limited to trusted entities (e.g., a clinic expert), preventing an adversary from viewing the system's outputs.}\deleted{Owners and intimate adversaries alike cannot access reports generated by \sys unless visiting a security clinic.} Additionally, only IPI-flagged time snippets would be kept, which are usually useless to the adversary, while helpful for the owner to assess the situation. Given these constraints, it is difficult for an intimate adversary to manipulate \sys to harm the owner.

\section{EVALUATIONS}\label{sec-5-eval}

\subsection{Dataset and Settings}
\label{subsec:dataset_settings}

\noindent\textbf{Data collection.} We collected data on two Android smartphones (Pixel 6A: Android 13 and Samsung A54: Android 14) using \sys, which continuously logs in the background without requiring privileged (root) access (Section~\ref{subsec:os_compatibility}). All related activities were approved by the Institutional Review Board (IRB).

\begin{table}[h!]
\centering
\caption{Common Apps and Their Categories}
\label{tab:apps}
\begin{tabular}{|l|l|}
\hline
\textbf{Category} & \textbf{App Name} \\
\hline
E-commerce & Amazon   \\
Communication & Gmail\\
Social Media & Instagram\\
Collaboration & Slack\\
Music & Spotify \\
Video Streaming & YouTube\\
\hline
\end{tabular}
\end{table} 

\begin{table*}[htbp!]
\centering
\caption{
Comparison of \sys~against baseline methods (end-to-end approaches and black box fusion approaches) for \REV{candidate IPI-risk event flagging across a 12-fold controlled evaluation.} Standard deviation errors are reported. Best scores per column are in \textbf{bold}. Second best scores per column are \underline{underlined}. 
}
\label{tab:fusion-results}
\begin{adjustbox}{width=\linewidth}
\begin{tabular}{llccccc}
\toprule
\textbf{Scope} & \textbf{Method} & \textbf{F1 Score $\uparrow$} & \textbf{FPR $\downarrow$} & \textbf{FNR $\downarrow$} & \textbf{Recall $\uparrow$} & \textbf{Precision $\uparrow$} \\
\midrule
\multirow{4}{*}{End-to-end Approaches} 
& Monolithic Supervised Model & \smallpm{0.663}{0.120} & \smallpm{0.346}{0.090} & \smallpm{0.288}{0.152} & \smallpm{0.712}{0.152} & \smallpm{0.633}{0.125} \\
& Anomaly Detection (AE-based) \cite{lin2020anomaly} & \smallpm{0.540}{0.166} & \smallpm{0.276}{0.135} & \smallpm{0.495}{0.173} & \smallpm{0.505}{0.173} & \smallpm{0.610}{0.162} \\
& User Authentication \cite{huh2023long} & \smallpm{0.908}{0.049} & \smallpm{0.181}{0.128} & \smallpm{\textbf{0.015}}{0.053} & \smallpm{\textbf{0.985}}{0.053} & \smallpm{0.845}{0.066} \\
& HAR & \smallpm{0.661}{0.088} & \smallpm{0.780}{0.058} & \smallpm{0.066}{0.046} & \smallpm{0.934}{0.046} & \smallpm{0.518}{0.098} \\
\cmidrule(lr){1-7}
\multirow{3}{*}{Black Box Fusion Approaches}
& MLP & \smallpm{0.828}{0.062} & \smallpm{0.103}{0.032} & \smallpm{0.141}{0.073} & \smallpm{0.827}{0.087} & \smallpm{0.834}{0.061} \\
& SVM & \smallpm{0.871}{0.048} & \smallpm{\underline{0.072}}{0.039} & \smallpm{0.161}{0.078} & \smallpm{0.839}{0.078} & \smallpm{0.911}{0.039} \\
& RF  & \smallpm{0.878}{0.053} & \smallpm{0.088}{0.062} & \smallpm{0.138}{0.058} & \smallpm{0.862}{0.058} & \smallpm{0.899}{0.069} \\
\cmidrule(lr){1-7}
\multirow{2}{*}{Ours}
& \sys w/o Context-Aware Fusion & \smallpm{\underline{0.920}}{0.055} & \smallpm{0.078}{0.098} & \smallpm{0.080}{0.069} & \smallpm{0.920}{0.069} & \smallpm{\textbf{0.924}}{0.067} \\
& \textbf{\sys w/ Context-Aware Fusion} & \smallpm{\textbf{0.928}}{0.034} & \smallpm{\textbf{0.070}}{0.034} & \smallpm{\underline{0.064}}{0.047} & \smallpm{\underline{0.936}}{0.047} & \smallpm{\underline{0.921}}{0.033} \\
\bottomrule
\end{tabular}
\end{adjustbox}
\end{table*}

\begin{table}[t!]
\centering
\caption{\REV{Evaluation of top-$k$ behavior inclusion on candidate IPI-risk event flagging performance}}
\label{tab:e2etopk}
\begin{tabular}{lccc}
\toprule
\textbf{Top-$k$} & \textbf{F1 Score $\uparrow$} & \textbf{FPR $\downarrow$} & \textbf{Recall $\uparrow$} \\
\midrule
Top-1 & \smallpm{0.928}{0.034} & \smallpm{0.070}{0.034} & \smallpm{0.936}{0.047} \\
Top-2 & \smallpm{0.935}{0.049} & \smallpm{0.022}{0.016} & \smallpm{0.902}{0.079} \\
Top-3 & \smallpm{0.981}{0.032} & \smallpm{0.016}{0.014} & \smallpm{0.981}{0.060} \\
\bottomrule
\end{tabular}
\end{table}

Participants completed a randomized list of 44 tasks (full list in Appendix~\ref{appendix:full_action_list}) designed to span both potentially IPI-related actions (e.g., modifying account settings) and neutral (e.g., watching videos, browsing items) across six common applications (Table \ref{tab:apps}). Before each task, an experimenter selected the corresponding label, and participants completed the task in their preferred manner. 

\noindent
\textbf{\INTERVIEW{Rationale for controlled simulation with healthy close pairs.}}
\INTERVIEW{Directly collecting ground truth from ongoing IPV cases would risk asking vulnerable people to produce, relive, or disclose harmful episodes, while still not yielding clean labels for intent or harm. As E2 emphasized in our formative interviews (Section~\ref{subsec:expert-interviews}), \textit{``[ground truth] in this situation is extremely hard ...''} and \textit{``you cannot just recruit abusers... that makes it so hard to study the real abuser and survivor pairs''} (E5). We therefore designed a controlled study as a best-effort technical simulation, similar to how vehicle-safety research uses simulated crash tests to measure whether sensors and frames respond to crash-like forces without causing real harm.}

\INTERVIEW{This choice is consistent with E1's description of scenario-based testing: although it is \textit{``quite removed from a real abuser,''} it can still \textit{``provide reasonable confidence on whether your system is doing, say, detection is working as intended.''} Healthy close pairs are not treated as survivor--adversary substitutes. Rather, the identity--action split (Section~\ref{subsec:formulating}) is precisely what allows ethical evaluation: it lets us test signal feasibility on the two observable components---non-owner-like use and IPI-relevant phone actions---while leaving survivor-facing utility, deployment safety, and clinic fit as limitations for future work (Section~\ref{subsec:limitation}). As E4 cautioned, in-lab testing \textit{``does not fully simulate what is going on in the real world,''} and we revisit this gap in our limitations.}

\label{subsection: participants}
\noindent\textbf{\REPRESENT{Participant recruitment and dataset rationale.}}
\REPRESENT{Participants were recruited through institution-affiliated email lists, word of mouth, and flyers. Interested participants completed an online screening questionnaire, and eligible participants were adults who regularly use smartphones and common mobile applications such as YouTube, Instagram, Amazon, Slack, and Spotify. The final cohort included 27 participants aged 19--30. Because \sys~operates on device-side interaction traces rather than self-reported attitudes or demographic perceptions, the relevant sampling goal for this feasibility study is to cover diverse phone-side usage patterns, execution paths, and owner/non-owner interaction traces. We therefore interpret participant representativeness at the level of observable smartphone operations, while leaving population-level deployment with broader survivor-centered cohorts to future work.}

\noindent\REPRESENT{(1) Ethical task-based evaluation. Directly involving people currently experiencing IPV in an early-stage phone-monitoring prototype raises substantial safety and ethical concerns for both participants and researchers. Following prior work~\cite{bellini2023digital,daffalla2023account,cai2024famos} and our IRB-approved protocol, we recruited adult volunteers with close relationships (e.g., couples, friends, roommates) to perform controlled phone-side actions derived from our IPI-risk action taxonomy. The protocol asks participants to complete concrete smartphone tasks, rather than to enact abuse, coercion, fear, or malicious intent. Thus, this study evaluates whether phone-side signals can support candidate IPI-risk event flagging, not whether \sys~can infer relationship status, malicious intent, or real-world IPV dynamics from phone traces alone.}

\noindent(2) Open-path task execution. Our taxonomy (Table~\ref{tab:actionsubaction}) specifies the target phone-side actions but imposes no fixed execution path. For example, a participant can complete a ``modify password'' task through settings, search, or an app-specific account page, producing natural variation in interaction traces. Participants also performed neutral activities outside the taxonomy, labeled \textit{NIO}, to cover risk-neutral smartphone use and \REPRESENT{reduce overfitting to only suspicious actions.} \REV{This design avoids training on a single scripted UI path and better matches how the same phone-side action may be completed through different menus, searches, or app-specific flows in practice.}

\noindent(3) App and action diversity. \REPRESENT{Participants completed 44 randomized tasks spanning IPI-risk actions and neutral activities across six common applications from distinct categories (Table~\ref{tab:apps}). This design exposes \sys~to different app interfaces, action semantics, and foreground-app contexts. In addition, our leave-one-app-out evaluation (Appendix~\ref{appendix:app_held_out}) shows less than 5\% performance drop, suggesting that \sys~is not tied to a single app interface and motivating future evaluation on a larger app set.}

\noindent(4) User and pair diversity. The dataset contains 27 distinct participants, including pre-identified close pairs. All evaluations use strict user-level non-overlapping train/test splits: in each leave-two-user-out fold, the entire owner--non-owner test pair is withheld from pretraining, and evaluation is performed only on this held-out pair. Six folds use genuine close pairs from the dataset, while the remaining folds use synthetic unrelated pairs to test broader non-owner generalization. \REPRESENT{This protocol supports cross-user evaluation within the sampled smartphone-proficient cohort and approximates the owner/non-owner access structure in our target scenario.}

\noindent\textbf{Dataset descriptions.} We collected the modalities listed in Table~\ref{tab:modalitycontent} using standard Android APIs without requiring root access (Section~\ref{subsec:os_compatibility}). The final dataset includes usage data from 27 participants (18 male, 9 female), including 9 pre-identified pairs (e.g., couples or close friends). On average, we collected 41.5 minutes from each participant, resulting in 18.7 hours of total data.

\noindent\textbf{Training-test split.} All evaluations follow a \underline{user-level leave-two-out} protocol \REV{designed to approximate the phone-access structure of our target scenario, with one participant treated as the device owner and another as a non-owner user}, as formulated in Section \ref{subsection: threatmodel}. We perform 12-fold cross-validation.
\noindent\textbf{Dataset processing.}  To process collected data, we first apply min-max normalization to the raw data streams, scaling each feature to the range [0, 1]. Data streams are collected at the highest available sampling rate of up to 130 Hz. To analyze the impact of lower sampling rates, we downsample the data to various frequencies: [1 Hz, 2 Hz, 5 Hz, 10 Hz, 20 Hz]. Next, we apply a sliding window to extract windows of inputs into our models. To evaluate the effect of different time windows on identity-branch's performance, we experimented with varying window durations: [1~s, 2~s, 5~s, 10~s, 20~s].

\noindent

\noindent1) \underline{Test user selection}: Select one user-pair as the test set, with one fixed as owner and the other as adversary. Six folds use the six\footnote{Three additional genuine pairs are excluded because their data were collected on a different phone model, violating the single-device assumption and our threat-model assumption that attackers attempt to access the same phone that the owner uses.} genuine couples/friends in our dataset; the other six folds use synthetic pairs formed by randomly selecting two unrelated participants.

\noindent2) \underline{Data partitioning}: The entire chosen user-pair for testing is withheld from the pretraining set. The owner's first five minutes are added to the fine-tuning set for on-device adaptation, as described in Section \ref{subsection:adaptprocess}. Three randomly selected users from the pretraining set are also moved to the fine-tuning set. Then the remaining 22 participants constitute the final pretraining data.

\noindent3) \underline{Model training and evaluation}: We pretrain on the 22-user dataset, fine-tune on the 3-user + owner-5-min-snippet set, and \textbf{\textit{evaluate the trained model solely on the held-out owner-adversary-pair}}. Unless specified, all results in Sections \ref{subection:e2eeval}--\ref{subection:l2eval} are averaged over these 12 folds and therefore reflect performance on completely unseen owners and adversaries. 

\REV{The end-to-end candidate-event label is determined by the rule: if the user is the owner, all samples are labeled as non-flagged; if the user is a non-owner and the action is NIO, the sample is labeled as non-flagged; otherwise, a non-owner performing an IPI-risk action is labeled as a candidate IPI-risk event.}

\noindent\textbf{Baseline methods.}
We group baselines into two families:

\noindent \underline{1. End-to-end Approaches}: (1) An LSTM-CNN-based monolithic supervised model that is trained directly on the IPI-event candidate labels; (2) Standalone Auto-Encoder-based anomaly detection \cite{lin2020anomaly} using reconstruction-error. (3) User Authentication: using our identity-branch as the sole detector (inspired by \cite{huh2023long}). It classifies a sample as IPI if the embeddings from the identity branch deviate significantly from the owner profile, based on reconstruction error and classification threshold (fixed to 0.5). (4) HAR: Since there is no comparable HAR baseline in the current literature, we leverage only our behavior-branch for this baseline to detect user activities. In this case, if the predicted class belongs to the IPI-related behavior set, the sample is labeled as an IPI-related behavior; else it is NIO.

\noindent \underline{2. Black Box Fusion Approaches}: \sys fuses the outputs of the identity and behavior branches in an interpretable and context-aware manner (Sections~\ref{subsec:formulating} and~\ref{scalable}). In the current machine learning landscape, the first approach that a developer would likely implement is to train a black-box predictor that concatenates the hidden states from both branches as input. To demonstrate the performance improvement of \sys's context-aware fusion, we evaluate against a black-box predictor implemented with an MLP, SVM, and Random Forest that uses the concatenated hidden vectors from the identity and behavior branches as input.

\noindent\textbf{Evaluation metrics.} According to our problem settings in Section \ref{subsec:formulating}, IPI-risk event flagging is formulated as a binary-classification task, so we report standard classification metrics including F1 score, precision, recall, false positive rate (FPR), false negative rate (FNR), and accuracy. \TOPK{Because the final flag is derived from the behavior branch's ranked action candidates, Top-1 is our primary setting; in Top-$k$ variants, $k$ controls how many behavior candidates are retained before the final binary flag. All event-level metrics are computed on the final binary output}. F1 summarizes the balance between precision and recall. FPR is particularly important in our context, as false alarms may trigger unnecessary interventions and erode trust in detection results. FNR captures missed candidate events, which could omit potentially relevant phone-side records from later review. Accuracy reflects overall correctness across classes. Full training details are provided in Appendix \ref{appendix:eval_settings}.

\begin{table*}[htbp!]
    \centering
    \caption{Comparison between \sys and baseline models for User Identification. Standard deviation errors are reported.}
    \label{tab:baseline}
    \begin{adjustbox}{width=\textwidth}
    \begin{tabular}{lccc ccc ccc c}
    \toprule
    \textbf{Method} 
    & \multicolumn{3}{c}{\textbf{Partner}} 
    & \multicolumn{3}{c}{\textbf{Stranger}} 
    & \multicolumn{3}{c}{\textbf{Overall}} 
    & \textbf{Model Size} \\
    \cmidrule(lr){2-4} \cmidrule(lr){5-7} \cmidrule(lr){8-10}
    & F1 Score $\uparrow$ & FPR $\downarrow$ & FNR $\downarrow$
    & F1 Score $\uparrow$ & FPR $\downarrow$ & FNR $\downarrow$
    & F1 Score $\uparrow$ & FPR $\downarrow$ & FNR $\downarrow$
    & \\
    \midrule
    KedyAuth 
    & \smallpm{0.864}{0.069} & \smallpm{0.413}{0.316} & \smallpm{0.014}{0.026}
    & \smallpm{0.901}{0.100} & \smallpm{0.214}{0.163} & \textbf{\smallpm{0.000}{0.000}}
    & \smallpm{0.883}{0.088} & \smallpm{0.314}{0.271} & \textbf{\smallpm{0.007}{0.019}}
    & 5,373~KB \\
    AuthentiSense 
    & \smallpm{0.880}{0.128} & \smallpm{0.057}{0.112} & \smallpm{0.169}{0.193}
    & \textbf{\smallpm{0.989}{0.015}} & \smallpm{0.022}{0.030} & \smallpm{0.004}{0.009}
    & \smallpm{0.935}{0.106} & \smallpm{0.039}{0.084} & \smallpm{0.086}{0.159}
    & 1,186~KB \\
    Ours (w/o TCM) 
    & \smallpm{0.961}{0.025} & \smallpm{0.077}{0.058} & \smallpm{0.013}{0.030}
    & \smallpm{0.948}{0.039} & \smallpm{0.091}{0.059} & \smallpm{0.027}{0.059}
    & \smallpm{0.954}{0.033} & \smallpm{0.084}{0.059} & \smallpm{0.020}{0.047}
    & \textbf{96~KB} \\
    \textbf{Ours (w/ TCM)} 
    & \textbf{\smallpm{0.998}{0.002}} & \textbf{\smallpm{0.003}{0.004}} & \textbf{\smallpm{0.002}{0.002}}
    & \smallpm{0.977}{0.041} & \textbf{\smallpm{0.006}{0.007}} & \smallpm{0.035}{0.077}
    & \textbf{\smallpm{0.987}{0.031}} & \textbf{\smallpm{0.005}{0.006}} & \smallpm{0.018}{0.057}
    & \textbf{96~KB} \\
    \bottomrule
    \end{tabular}
    \end{adjustbox}
\end{table*}  

\subsection{End-to-End System Performance}

\label{subection:e2eeval}

The end-to-end IPI-risk event flagging performance is summarized in Table~\ref{tab:fusion-results}.  \sys~achieves the highest F1 score of 0.928, outperforming all baseline methods that attempt to flag candidate IPI-risk events directly. The same is true for other key metrics, such as Recall and FPR; in particular, the lower FPR indicates fewer false alarms, which is important in this setting because false alarms could lead to the owner's overreaction and unnecessary concerns. Compared to black-box fusion (e.g., MLP, SVM, RF), AID’s unified design with context-aware fusion offers superior robustness and lower variance. The ablation without this module leads to a notable performance drop (e.g., FNR increases from 0.064 to 0.080), validating its effectiveness. The results  are reported for action (9 classes) level behaviors, but the other two levels of behavior (5 classes and 28 classes) show similar trends.

Beyond raw accuracy, \sys’s context-aware fusion and risk analyzer ensures transparency and modularity, which contrasts with black-box meta-learners or monolithic models. While anomaly detection models struggle due to overlapping behavioral patterns between benign and malicious activities, \sys disentangles identity and behavior signals to provide a clearer and explainable decision boundary. 

AID's superior stability and lower false positive rate are further illustrated in a visual case study in Appendix~\ref{appendix:e2e_case}.

    \noindent\textbf{Providing comprehensive analysis via top-k behavior inclusion.}
    To mitigate behavior-level ambiguity from indirect signals, AID allows top-$k$ behavior predictions to support downstream risk estimation and interpretation. As shown in Table~\ref{tab:e2etopk}, incorporating top-2 or top-3 predictions significantly improves F1 score (from 0.928 to 0.981) and reduces false positives (FPR drops from 0.070 to 0.016). This highlights the value of a ranked contextual behavior interpretation, especially when actions are hard to distinguish from sensor data alone. \REV{The top-k output design supports a more comprehensive, human-in-the-loop review for future clinic-mediated use.}

\subsection{Effectiveness of User Identity Branch}\label{Level 1 eval}

Although \sys~is primarily for IPI-suspicious event flagging, its identity-branch alone could be used to effectively authenticate mobile users. To illustrate its performance, we compared our identity branch as a user-authentication tool with KedyAuth~\cite{huh2023long} and AuthentiSense~\cite{fereidooni2023authentisense}, two state-of-the-art user authentication frameworks for mobile devices, both of which use autoencoder-based anomaly detection with different classifier heads (CNN or feature-based SVM). We implemented and trained both methods using the same datasets and setting as \sys. 

Shown in Table~\ref{tab:baseline}, \sys~achieves higher F1, lower FPR and comparable FNR, and does so with a significantly smaller model size—highlighting its suitability for stealthy and resource-efficient deployment.
    
We further break down performance by attacker relationship. Both KedyAuth and AuthentiSense struggle in cases where a close partner is the non-owner due to behavioral similarity, yielding high FARs (e.g., 0.413 for KedyAuth). In contrast, AID maintains strong partner detection (F1 = 0.998, FAR = 0.003) without sacrificing performance against strangers. This indicates our model’s ability to handle fine-grained differences even in challenging IPI contexts. We observed that the embeddings generated by \sys compared to KedyAuth and AuthentiSense yielded greater distances and lower cosine similarity scores, which enables \sys to better distinguish between the phone's owner and non-owner (Appendix~\ref{appendix:user_id_encoder_embedding}).

We also explored the impacts of pretraining and fine-tuning (Appendix~\ref{appendix:pretrain_finetune}), sampling rates and window sizes (Appendix~\ref{appendix:vary_sample_rate_window_size}), number of LSTM heads (Appendix~\ref{appendix:num_lstm_heads}), and classification head choice (Appendix~\ref{appendix:id_classification_head_choice}).

\begin{figure}[t!]
    \centering
    \begin{minipage}{0.4\linewidth}
        \centering
        \includegraphics[width=\linewidth]{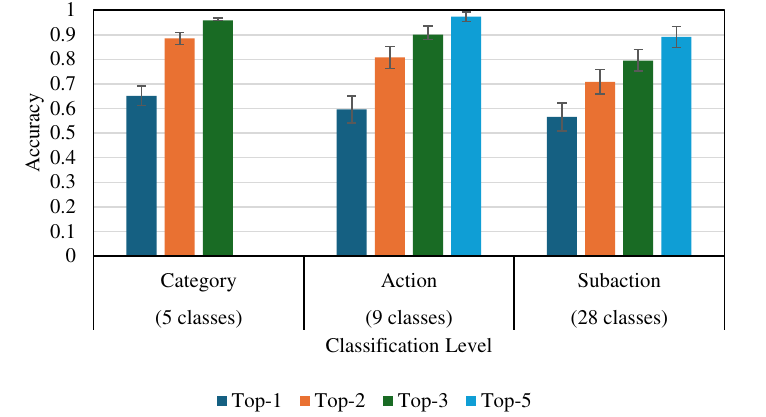}
    \end{minipage}
    \hspace{0.02\linewidth} 
    \begin{minipage}{0.4\linewidth}
        \centering
        \includegraphics[width=\linewidth]{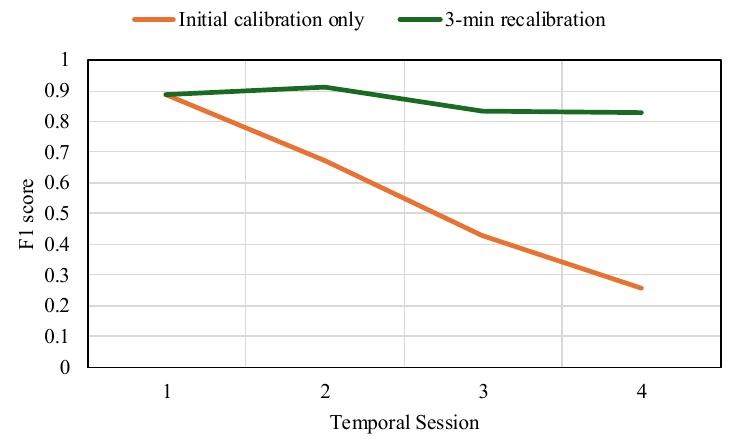}
    \end{minipage}
    
    \caption{(left) Top-k accuracy across granularity levels for IPI-related behavior classification. \LONG{(right) Averaged F1 trend across four longitudinal sessions}.}
    \label{fig:mod2perf}
\end{figure}

\subsection{Effectiveness of Behavior Branch}

\label{subection:l2eval}

\noindent\textbf{Classification accuracy.} On top of identity-branch, \sys's behavior-branch detects \ipvact at the three granularity levels according to our proposed taxonomy (Table~\ref{tab:actionsubaction}). The performance is summarized in Figure~\ref{fig:mod2perf} (left), where we report the top-k performance for k = 1, 2, 3, and 5 (e.g., the ground truth category, action, or subaction is in the list of top-k most probable behaviors). In the case of top-1, or logging only the most probable behavior, \sys performs significantly better than random guessing (accuracy of around 0.6 to 0.7 vs. random guessing accuracy of 0.2). This far from perfect performance illustrates the difficulty of distinguishing \ipvact, using only a limited set of data streams, to avoid discovery by adversaries.

However, by slightly broadening our logging to the top-k, \REV{the probability that the ground-truth label appears in the ranked behavior list increases substantially}. For instance, reporting the top-5 most probable \textit{subactions} yielded an accuracy of 0.891 and reporting the top-3 \textit{actions} boosted the accuracy to 0.900. These results highlight how certain \ipvact may exhibit similar patterns across multiple modalities, especially at finer-grained levels (e.g., subactions). This reinforces the need for systems like \sys to account for ambiguity by ranking behaviors, rather than solely relying on single-point predictions, which still offers expert reviewers actionable insights.

\noindent\textbf{Case study: behavior ambiguity and top-2 interpretation.} Table~\ref{tab:top2_confusion} highlights representative cases in which multiple behaviors exhibit high contextual similarity. For instance, \textit{viewing an account} is frequently confused with \textit{altering account settings}, and \textit{uploading content} often co-occurs with benign. Despite this inherent ambiguity, the correct label frequently appears within the top-2 predictions (e.g., 30.45\% of \textit{view account} cases have \textit{alter account settings} as top-2), with reasonably high average confidence scores. These results suggest that while fine-grained behavior classification remains challenging, \TOPK{top-k outputs provide a mechanism to expose uncertainty in user-interaction interpretation and support cautious downstream expert review}.

\noindent\textbf{Classification backbone comparisons.} We compare against LSTM, CNN, and Transformer backbones using identical input features and training settings. Our proposed LSTM-CNN hybrid architecture achieves the best top-k accuracy across most behavior levels.

Table~\ref{tab:mod2clf} shows the performance of \sys in classifying \ipvact with different backbone architectures. Our hybrid LSTM-CNN architecture which extracts both temporal (LSTM) and spatial (CNN) features, significantly outperforms single-type architectures. Across all scenarios this hybrid architecture also outperforms a transformer architecture, which generally outperforms other architectures when data are abundant and model sizes are large. However, in our scenario where models are small (only hundreds of thousands of parameters), transformer performance typically suffers, which matches our observations. While LSTM-CNN did not outperform a purely CNN or LSTM approach in determining the category (5-class) of behavior, the performance is essentially equal, with less than $1\%$ difference.

\subsection{\LONG{Temporal Longitudinal Study}}
\label{sec:temporal_longitudinal}

\LONG{In addition to the 27-user study, we conducted a small-scale longitudinal study with two new couples to examine \sys's stability across repeated use. Each couple completed four collection sessions, spanning 21 days for one couple and 14 days for the other. In each session, participants completed approximately half of the task list used in the main study, producing about 20 minutes of data per session on average and 5.25 hours of data in total. In the final session, one couple used the phone in a dark room and the other used the phone in a moving vehicle, providing an initial stress test under environmental changes.}

\LONG{We evaluate Top-1 end-to-end IPI-risk event flagging and compare \sys~with an ablated protocol that uses a model, calibrated and tuned only on data from the first session, across all sessions. We compare this against a recalibration protocol, where we use the first 3 minutes of owner-confirmed data in each session for lightweight session-level recalibration, and evaluate on the remaining owner data together with the non-owner data from the same session.}

\LONG{Table~\ref{tab:temporal_longitudinal} reports results averaged over the two couples and four sessions. Without recalibration, \sys~shows substantial temporal drift, with mean F1 of 0.562 and high variance. With 3-minute recalibration, mean F1 improves to 0.865, FAR decreases from 0.461 to 0.146, and FSR decreases from 0.338 to 0.101. Figure~\ref{fig:mod2perf} (right) further shows that the fixed-profile protocol degrades across sessions, while the recalibrated protocol remains comparatively stable, including in the final environmentally shifted session. These results suggest that temporal user-behavior drift is a practical concern, but a short recalibration step can substantially improve longitudinal robustness. Appendix~\ref{app:contamination} further presents a controlled contamination-and-recovery analysis, where one intermediate recalibration buffer is mixed with non-owner samples and later sessions resume clean owner-data recalibration, showing that AID remains relatively stable under isolated contamination.} 

\begin{table}[t]
\centering
\caption{\LONG{Temporal longitudinal evaluation under the Top-1 end-to-end IPI-risk event flagging rule. Results are averaged over two couples and four sessions.}}
\label{tab:temporal_longitudinal}
\small
\begin{tabular}{lcccc}
\toprule
Protocol & F1 & FAR & FSR & Min session F1 \\
\midrule
Initial calibration only 
& $0.562 \pm 0.299$ 
& 0.461 
& 0.338 
& 0.260 \\
3-min session-level recalibration 
& $0.865 \pm 0.078$ 
& 0.146 
& 0.101 
& 0.827 \\
\bottomrule
\end{tabular}
\end{table}

\begin{table}[]
\small
\centering\caption{Top-1--3 accuracy (mean ± std) of different model architectures across three behavior classification levels: Category (5 classes), Action (9 classes), and Subaction (28 classes). Best performance per row is \textbf{bolded}.}
\label{tab:mod2clf}
\begin{tabular}{llcccc}
\toprule
\textbf{\shortstack{Classification\\Level}} & \textbf{\shortstack{Accuracy\\~}} & \textbf{\shortstack{LSTM\\~}} & \textbf{\shortstack{CNN\\~}} & \textbf{\shortstack{Transformer\\~}} & \textbf{\shortstack{LSTM-CNN\\(Ours)}} \\
\midrule
\multirow{3}{*}{\shortstack{Category\\(5 classes)}} 
& Top-3 & \smallpm{0.944}{0.0174} & \smallpm{0.950}{0.0172} & \smallpm{0.939}{0.0207} & \textbf{\smallpm{0.959}{0.0104}} \\
& Top-2 & \smallpm{0.860}{0.0276} & \smallpm{0.866}{0.0277} & \textbf{\smallpm{0.919}{0.0263}} & \smallpm{0.885}{0.0249} \\
& Top-1 & \smallpm{0.626}{0.0491} & \textbf{\smallpm{0.671}{0.0345}} & \smallpm{0.613}{0.0515} & \smallpm{0.652}{0.0390} \\
\midrule
\multirow{3}{*}{\shortstack{Action\\(9 classes)}} 
& Top-3 & \smallpm{0.863}{0.0469} & \smallpm{0.882}{0.0384} & \smallpm{0.865}{0.0432} & \textbf{\smallpm{0.900}{0.0354}} \\
& Top-2 & \smallpm{0.766}{0.0478} & \smallpm{0.784}{0.0462} & \textbf{\smallpm{0.813}{0.0399}} & \smallpm{0.808}{0.0446} \\
& Top-1 & \smallpm{0.573}{0.0545} & \textbf{\smallpm{0.630}{0.0511}} & \smallpm{0.568}{0.0527} & \smallpm{0.596}{0.0542} \\
\midrule
\multirow{3}{*}{\shortstack{Subaction\\(28 classes)}} 
& Top-3 & \smallpm{0.705}{0.0773} & \smallpm{0.757}{0.0503} & \smallpm{0.722}{0.0645} & \textbf{\smallpm{0.795}{0.0458}} \\
& Top-2 & \smallpm{0.619}{0.0767} & \smallpm{0.667}{0.0521} & \smallpm{0.667}{0.0647} & \textbf{\smallpm{0.709}{0.0496}} \\
& Top-1 & \smallpm{0.465}{0.0830} & \smallpm{0.551}{0.0602} & \smallpm{0.483}{0.0650} & \textbf{\smallpm{0.566}{0.0575}} \\
\bottomrule
\end{tabular}
\end{table}

\begin{table}[]
\centering
\caption{Representative cases highlighting ambiguity in contextual behavior classification.}
\label{tab:top2_confusion}
\begin{tabular}{|c|c|c|c|}
\hline
\textbf{True Label} & \textbf{Prediction (Top-1 / Top-2)} & \textbf{Percentage} & \textbf{Avg. Score (Top-1 / Top-2)} \\
\hline
Send content & send content / alter account settings & 15.97 & 0.63 / 0.21 \\
\hline
Upload content & upload content / NIO & 27.51 & 0.65 / 0.21 \\
\hline
View account & view account / alter account settings & 30.45 & 0.51 / 0.27 \\
\hline
Alter account settings & alter account settings / view account & 25.59 & 0.56 / 0.26 \\
\hline
Modify content & modify content / alter account settings & 16.25 & 0.49 / 0.27 \\
\hline
NIO & NIO / view account & 15.69 & 0.57 / 0.21 \\
\hline
\end{tabular}
\end{table}

\section{DISCUSSION AND FUTURE WORK}\label{sec-6-discussion}

\subsection{Limitations}
\label{subsec:limitation}

\noindent\textbf{\REV{Ecological validity and staged validation.}} \REV{A primary limitation of this work is that our evaluation does not observe real IPV episodes or measure survivor outcomes. The controlled study cannot reproduce fear, coercion, strategic adaptation, malicious intent, or the broader power dynamics of an abusive relationship. This limitation is deliberate: our expert consultations emphasized that real IPV ground truth is difficult and ethically risky to obtain in an early-stage phone-monitoring prototype, and that a technical system should not infer abuse or intent from phone traces alone. Therefore, we do not treat non-survivor participants as substitutes for survivors or intimate adversaries. Instead, our study evaluates a narrower construct: whether real smartphone users performing concrete phone-side operations from an IPI-derived taxonomy generate signals that can support candidate IPI-risk event flagging.}

\REV{We partially mitigate this limitation through staged validation. The taxonomy is grounded in prior IPV literature; participants performed tasks on real phones and common applications; task execution was open-path rather than scripted; the evaluation used strict user-level held-out pairs, including genuine close pairs; and the longitudinal study introduced repeated sessions plus environmental shifts such as a dark room and a moving vehicle. These choices support ecological validity at the level of observable phone-side operations, not at the level of lived IPV dynamics. Future work should move from this system-feasibility stage to clinic-mediated field evaluation with professional oversight, informed consent, withdrawal and deletion procedures, and no automated alerts or interventions.}

\noindent\textbf{\FINALREV{Impact of real relationship dynamics on system effectiveness.}}
\FINALREV{
Real relationship dynamics may affect \sys at several levels.
First, close partners may develop similar daily routines and smartphone-use habits, making owner/non-owner identification more difficult; an aware partner may also deliberately imitate the owner.
This challenge is reflected in the two continuous-authentication baselines in Table~\ref{tab:baseline}, which show lower identification performance and weaker separation for close partners than for strangers.
Although the identity branch of \sys achieved strong separation among the healthy close pairs in our controlled study, longer-term shared routines and deliberate imitation were not evaluated and remain important steps for future work.}

\FINALREV{Second, \sys flags a candidate event only when non-owner-like use is combined with an IPI-related phone-side action. This rule does not cover cases in which the owner performs an action under pressure from a partner: the user remains owner-like even though the action may be harmful in its relationship context. Additionally, consensual device and credential sharing is common in close relationships~\cite{Dragiewicz2026smartphone,rogers2023technology}. For example, a non-owner may change a password or account setting at the owner's request for a benign purpose. Such an event may be correctly recognized at both the identity and action levels but still be benign or unclear in its relationship context. Because \sys cannot observe consent, coercion, or relationship history, its outputs should remain candidate records for later review by trained professionals rather than determinations of abuse or harmful intent (Section~\ref{subsec:flagging}).}

\FINALREV{Third, fear, stress, or hurried use may change how the owner holds and operates the phone, including posture, interaction pace, and application-switching patterns. These changes may affect IMU, system, and interaction signals, but their effects on \sys remain unknown. Real relationship dynamics may also limit when and by whom recalibration is performed, allowing partner-generated or forced-use data to enter the owner calibration set. Our controlled longitudinal and contamination analyses (Appendix~\ref{app:contamination}) show sensitivity to across-session changes and to one isolated contamination condition, but repeated, covert, or coercive contamination remains untested.
Finally, an aware partner may change the application, timing, or action path to avoid detection; our open-path tasks capture ordinary execution differences, but not goal-directed evasion. These are all important dimensions that need careful evaluation and design to address in future work.
}

\label{risk_discussion}
\add{\noindent\textbf{On the risk of discovery and escalation.} \REV{\sys is designed for early concern scenarios in which a device owner collects structured phone-side records to support informed decision-making in consultation with professionals} (Section~\ref{sec-3-threatModel}). However, it is practical to consider the inherent risk and possible conflict escalation after discovery by the intimate adversaries. We acknowledge that no perfect technical tool, including \sys, could eliminate this risk.}

\add{One of our design philosophies is that \sys~leads to less severe consequences upon discovery compared to existing tools. For instance, unlike conventional continuous authentication systems, \sys~runs silently only for evidence gathering and does not send an immediate alert when detecting risks, which reduces confrontations between owners and adversaries. Additionally, drawing on prior works and designs, we strongly recommend that users engage in \textbf{rigorous safety planning}~\cite{freed2018stalker, havron2019clinical, freed2019my} with clinical security experts, ahead of real use. Users are also advised not to make any changes or moves upon acquiring detection results before meticulous consultation with security experts. Ultimately, in life-threatening situations or cases involving a high-tech adversary (e.g., one who can root the device and obtain superuser privileges), users should seek professional support and law enforcement~\cite{koshelev2025investigating}, which standalone technical tools such as \sys (Section~\ref{sec:stealthy_functioning}) cannot replace.}

\add{Lastly, while security experts are actively developing specialized security techniques which we believe~\sys could function alongside, they are complementary to our core scientific contribution of establishing a new detection methodology. Therefore, we view their future integration as a promising direction to further enhance \sys's resilience to adversary discovery. For instance, by adopting asymmetric keys, the device only has a public key from a trusted entity to encrypt the data while only the entity has the private key to decrypt the data for analysis, which would make \sys more tamper-evident. Additionally, app-hiding frameworks like HideMyApp~\cite{pham2019hidemyapp} could further hide its presence on the device.}

\noindent\textbf{\COMPATIBLE{Evaluation platform.}} \COMPATIBLE{We note that AID is currently a proof-of-concept Android implementation whose operability and stealth depend on Android-specific Accessibility Service, background-service, and permission semantics; therefore, its visibility on newer Android versions and customized Android distributions may vary, and extension to non-Android platforms such as iOS, PCs, or routers would require platform-specific redesign of sensing interfaces, permission models, and deployment workflows rather than direct portability.}

\subsection{Future Research Direction}
\label{subsec:future}

During the development of our methodology and evaluations, we have identified the following viable future directions.

\noindent\textbf{\FINALREV{Bridging controlled and real-world participation.}} Existing evaluations have verified \sys's feasibility as a proof-of-concept system. 
\FINALREV{Here we outline a gradual path needed to translate 
\sys from controlled studies to the real-world. First, the system workflow and report design should be reviewed with a wider range of IPV advocates, clinic staff, and survivor representatives who are willing and able to participate safely. This early stage can use fictional scenarios and mock \sys reports without collecting new phone data. Such work can help decide which report contents are useful, how uncertainty should be shown, who should be able to access the records, what options the report should provide, and how participants can pause, withdraw, or request data deletion. Second, future studies can broaden the controlled evaluation in two ways. Larger and longer studies with healthy close pairs can move from assigned tasks and lab-provided phones toward participant-controlled devices and more natural phone use. Separately, non-owner participants can be given a clear goal to imitate the owner or avoid detection by changing the application, timing, interaction pattern, or action path. These studies can test more realistic use conditions and deliberate evasion, but they still cannot reproduce abusive intent.}

\FINALREV{Only after these lower-risk studies and a renewed safety and ethics review should a small, voluntary, clinic-mediated, and participant-controlled background feasibility study with people affected by IPV be considered. Such a study should be carefully reviewed, keep participants in control throughout, collect records without automatically taking action, and include designed fallback plans in case safety concerns arise. Before any larger or longer-term field deployment, the study should undergo another safety and ethics review.}

\add{\noindent\textbf{Expansion of coverage of apps and behavior types.} In this work, six apps from six representative categories are selected for data collection and evaluation, demonstrating \sys's applicability across diverse usage scenarios; our IPI behavior taxonomy, using the combination of the \textit{Int} and the \textit{App} modalities, enabled effective behavior classification. However, there remains room for improvement in both granularity and accuracy for behavior detection, especially for wider usage cases. This is due to the low-fidelity nature of the available modalities we found and the ambiguity observed across certain behavior classes. \LONG{Future work could expand AID from behavior flagging to richer evidence characterization by combining its identity and action signals with complementary package analysis, permission analysis, content analysis, and network analysis, such as anti-spyware scanning, location access auditing, and harmful content detection, to support more detailed review of stalkerware, tracking apps, impersonation, and other IPI risk activities in clinic mediated settings}.}

\section{CONCLUSION}\label{sec-8-conclusion}

\add{This work addresses the challenge of systematically flagging evidence of Intimate Partner Infiltration (IPI) on a smartphone by formulating it as a problem of jointly analyzing unauthorized user identity and potentially risky behaviors.} \deleted{In this work, we formalized the threat model of Intimate Partner Infiltration (IPI) through a custom-designed taxonomy, and} \add{We} presented \sys, a \add{mobile system for} automated IPI detection that \deleted{jointly analyzes user identity and behavior} \add{adopts} \deleted{via} a dual-branch architecture that realizes this joint-reasoning model using OS-level and physical signals on mobile devices -- under stealth, safety, and privacy constraints.  \TOPK{Our 27-participant evaluation shows that \sys~flags candidate IPI-risk events with 0.928 F1 and 7.0\% FPR under the Top-1 classification. Retaining the top-$k$ candidate behaviors further improves event flagging to 0.981 F1 and 1.6\% FPR, while preserving the ambiguity for later human review}. \add{To conclude, this work establishes the feasibility of a new and automated paradigm for evidence gathering in the IPI context, offering a powerful and scalable tool that complements existing solutions like security clinics in supporting victims.}

\newpage
\bibliographystyle{ACM-Reference-Format}
\bibliography{paper}

\newpage

\appendix

\section{ETHICS CONSIDERATIONS}
\label{appendix-ethics}

    \noindent\textbf{Institutional Review Board (IRB) approval.} This study was reviewed and approved by the Institutional Review Board (IRB) at our institution.
    
    \noindent\textbf{Participant selection and psychological safety.} Our study did not involve direct interaction with real IPV abusers or victims. Instead, we recruited couples and friends in healthy relationships. This approach was chosen to avoid causing psychological or physical harm to vulnerable populations while still collecting data with behavioral similarities valuable for the development of IPV detection systems. By doing so, we ensured the safety and well-being of participants.
    
    \noindent\textbf{Use of deception and debriefing.} At the beginning of the data collection process, we employed partial disclosure by framing the study as a ``Smartphone Usage Analysis.'' This deliberate use of deception was necessary to prevent participants from altering their behavior due to the knowledge of the study’s association with IPV-related research, which could bias the data collection and affect the reliability of general behavioral patterns.
    
    After the data collection phase concluded, participants were fully debriefed. During the debriefing, we disclosed the true purpose of the study, explained why deception was necessary, and provided them with an opportunity to ask questions and voice any concerns.
    
    Importantly, participants were given the right to withdraw their data at any point after the debriefing if they felt uncomfortable with the study or its purpose. This ensured that their autonomy and rights were respected throughout the process. \add{During our collection, no participant decided to withdraw after debriefing.}
    
    \noindent\textbf{Privacy and anonymity.} To protect participants’ privacy, we designed the study to inherently anonymize all collected data. No personal or identifiable information was logged during the process. Participants used lab-provided smartphones and lab-provided app accounts, ensuring minimal risk of privacy leakage.
    
    Additionally, the collected data were securely transferred and stored on an encrypted lab server accessible only to authorized personnel. These measures ensured that the risk of privacy breaches was effectively mitigated.

\section{\INTERVIEW{INTERVIEW QUESTIONS}}
\label{app:interview-questions}

\INTERVIEW{Table~\ref{tab:interview-protocol} presents our semi-structured interview protocol, organized into four sections that move from broad context to concrete design and evaluation considerations. Indented sub-questions were used as follow-up probes at the interviewer's discretion.}

\begin{table}[t]
\centering
\caption{\INTERVIEW{Semi-Structured Interview Protocol}}
\label{tab:interview-protocol}
\renewcommand{\arraystretch}{1.2}
\footnotesize
\begin{tabularx}{\linewidth}{@{} p{0.7cm} X @{}}
\toprule
\multicolumn{2}{@{}l}{\textit{\textbf{\INTERVIEW{Context}}}} \\
\midrule
\INTERVIEW{Q1} & \INTERVIEW{What do you think are the biggest differences between technology-facilitated IPV/IPI cases and more traditional cybersecurity incidents?} \\
\INTERVIEW{Q2} & \INTERVIEW{What are the biggest challenges in practice when trying to support real victims and survivors in these situations?} \\
\midrule
\multicolumn{2}{@{}l}{\textit{\textbf{\INTERVIEW{Technology in IPV}}}} \\
\midrule
\INTERVIEW{Q3} & \INTERVIEW{What current technical tools or practices are being used to help notice, verify, or respond to concerning device or account activity in IPV-related situations?} \\
   & \INTERVIEW{\quad-- Which kinds of signals seem most actionable in practice?} \\
   & \INTERVIEW{\quad-- What limitations do these tools have in practice?} \\
\midrule
\multicolumn{2}{@{}l}{\textit{\textbf{\INTERVIEW{Requirements for Technology Solutions}}}} \\
\midrule
\INTERVIEW{Q4} & \INTERVIEW{What characteristics are important for a technology solution intended to help in IPV/IPI-related situations?} \\
   & \INTERVIEW{\quad-- If you were to design a helping tool, what would be your design choices?} \\
\INTERVIEW{Q5} & \INTERVIEW{If a tool were designed to flag concerning actions or suspicious activity potentially related to IPV, who could it realistically help, and in what situations?} \\
\midrule
\multicolumn{2}{@{}l}{\textit{\textbf{\INTERVIEW{Deployments and Validation}}}} \\
\midrule
\INTERVIEW{Q6} & \INTERVIEW{What would be a reasonable and responsible way to do early validation, and what kinds of approaches would you be cautious about?} \\
   & \INTERVIEW{\quad-- If none, is the concern mainly about ethics, realism, or interpretation?} \\
   & \INTERVIEW{\quad-- If this evaluation approach seems problematic, what would be a better substitute?} \\
\bottomrule
\end{tabularx}
\end{table}

\section{EVALUATION SETTINGS}
\label{appendix:eval_settings}
We train and evaluate the models on a Linux server running on Ubuntu 22.04.5 LTS with an NVIDIA L40 GPU, using TensorFlow 2.17.0 and Python 3.12.7.

The AutoEncoder for \textit{user identification} is pretrained on a self-supervised reconstruction task for a maximum of 100 epochs with a batch size of 512, using the Adam optimizer and a learning rate of 1e-3. We incorporate early stopping with a patience of 5 epochs and a minimum improvement threshold of 0.0001. Mean Squared Error (MSE) is used as the loss function to reconstruct the input data. Similarly, the \textit{behavior classification} model is trained for 50 epochs with a batch size of 512, using the Adam optimizer and a learning rate of 1e-3, and is optimized with categorical cross-entropy as the loss function.

\begin{figure}[]
\centering\includegraphics[width=0.7\linewidth]{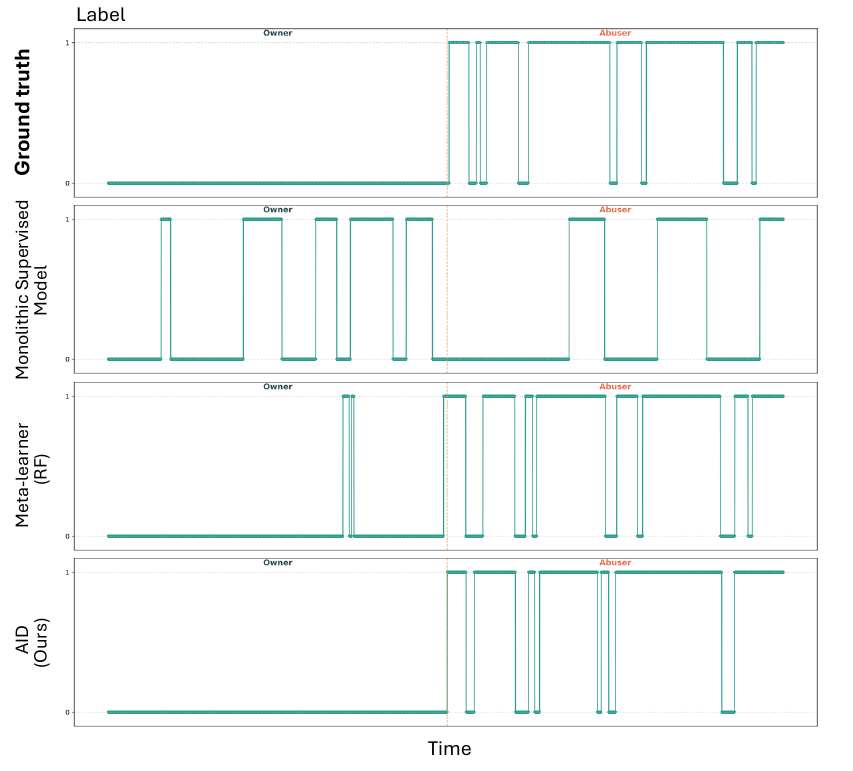}%
\caption{IPI detection outputs for a sample pair.}\label{fig:e2e-case}
\end{figure}

\section{CASE STUDY: VISUALIZING END-TO-END DETECTION PERFORMANCE}
\label{appendix:e2e_case}
Figure~\ref{fig:e2e-case} shows detection outputs for a sample owner-adversary pair. The monolithic model triggers frequent false positives during the owner phase (during which the owner is using the device), while the RF meta-learner remains less but notably noisy. In contrast, AID maintains stable predictions, with fewer false alarms before the transition and prompt adversary detection after. These results illustrate that \sys's two-stage approach determines the intermediate outputs (user ID and behavior) before fusing them, enabling \sys to better model and disentangle users and their behaviors. All baselines attempting to directly estimate IPI risk with a single end-to-end model could not successfully learn these characteristics.

\section{USER IDENTIFICATION ENCODER COMPARISON IN EMBEDDING SPACE}
\label{appendix:user_id_encoder_embedding}
\begin{table}[htbp!]
\centering
\caption{Embedding analysis based on Euclidean Distance (higher is better) and Cosine Similarity (lower is better) between partner and stranger samples.}
\label{tab:embedding-analysis}
\begin{tabular}{l|cc|cc}
\toprule
\textbf{} & \multicolumn{2}{c|}{\textbf{Euclidean Distance} $\uparrow$} & \multicolumn{2}{c}{\textbf{Cosine Similarity} $\downarrow$} \\
\textbf{Method} & Partner & Stranger & Partner & Stranger \\
\midrule
AuthentiSense & 0.405 & 0.422 & 0.913 & 0.904 \\
KedyAuth      & 0.791 & 0.843 & 0.672 & 0.633 \\
\textbf{\sys for user identification (Ours)} & \textbf{1.425} & \textbf{1.331} & \textbf{-0.032} & \textbf{0.098} \\
\bottomrule
\end{tabular}
\end{table}

        To understand the superior intimate partner detection performance of our user identification module, we analyze the embedding space produced by each model. As shown in Table~\ref{tab:embedding-analysis}, AID yields significantly greater Euclidean distances and lower cosine similarity between owner and attacker samples—particularly for partners. Notably, our embeddings achieve negative cosine similarity (\(-0.032\)), indicating strong directional separation even for behaviorally similar partners. In contrast, baseline methods cluster partners and owners closely (e.g., 0.913 in AuthentiSense), reducing their ability to detect social insiders. These results confirm that AID learns more discriminative and generalizable representations.

\section{EFFECTS OF PRETRAINING AND FINE-TUNING}
\label{appendix:pretrain_finetune}

\begin{figure}[htbp!]
\centering\includegraphics[width=0.6\linewidth]{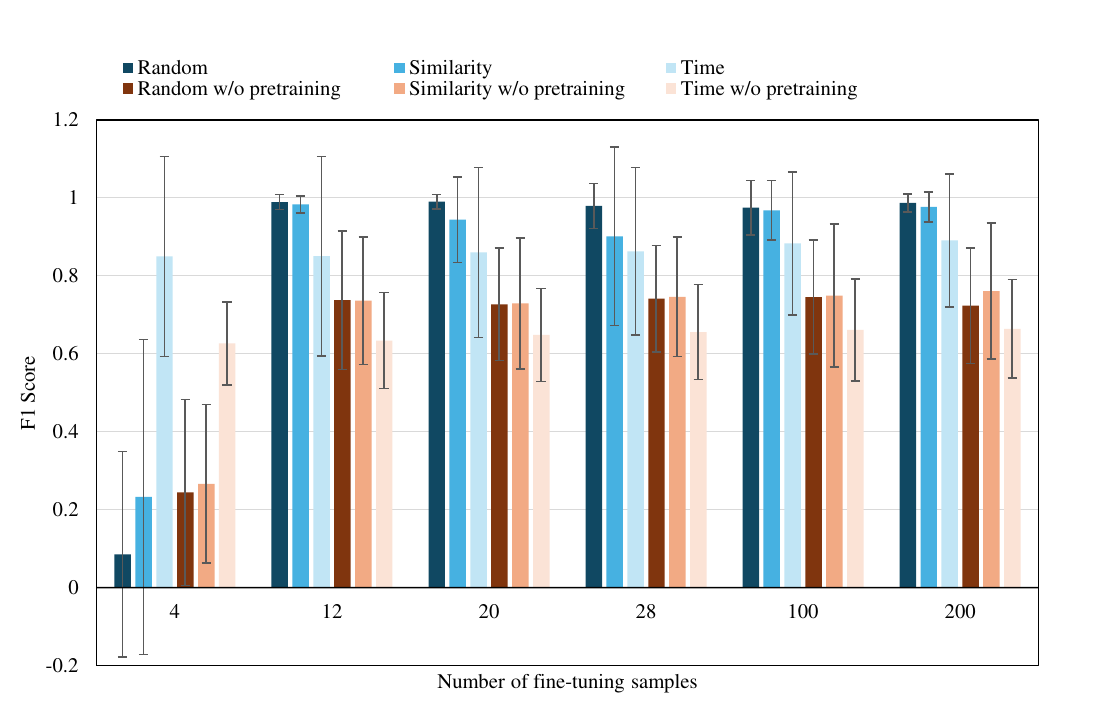}%
\caption{Impact of pretraining and fine-tuning scheme.}\label{fig:pretrain_effect}
\end{figure}

            Here, we analyze the impact of pretraining and fine-tuning. As shown in Figure~\ref{fig:pretrain_effect}, we vary the number of samples used to fine-tune the base model to the phone's owner (x-axis), which is collected from the user during a short one-time 5-minute calibration phase when the user initially installs \sys. We also ran experiments, with and without pretraining. Finally, we also looked at different selection schemes for choosing the windows used during fine-tuning. Ultimately, \sys adopts a random selection scheme (e.g., choose $n$ random windows of data provided by the user), which had the highest performance. We compared against selecting fine-tuning windows based on \textit{chronological time} and \textit{similarity}. 
        
        When selecting based on \textit{chronological time}, we simply use the first $n$ consecutive windows provided by the user. We believe that this scheme performed worse than random selection because the windows collected at a specific time are likely less diverse than randomly selecting windows across the entire calibration session, where users are likely performing a variety of different actions throughout.
        
        When selecting based on \textit{similarity}, we select windows from the owner and from other users that are similar to each other (measured by cosine similarity). The intuition is that it is more difficult to distinguish inputs that are more aligned, so incorporating them during the fine-tuning phase could potentially help the model distinguish these harder cases. To achieve a balance, we select windows in a specific ratio: 30\% hard (most similar), 50\% mid (moderately similar), and 20\% easy (least similar). However, these cases are likely not completely indicative of the general behavior of the owner and would require more examples to generalize well. Randomly sampling provides a high chance of selecting diverse samples, including both samples that embody a user's typical behavior and samples that may be difficult to distinguish from other users.
    
        As shown in Figure~\ref{fig:pretrain_effect}, the peak F1 score is achieved with our random selection method using 20 windows for fine-tuning and remains consistently high as the number of windows increases. In comparison, similarity-based selection exhibits a similar trend to that of the random method but plateaus at a lower F1 score. Furthermore, it shows instability as the training set size changes, fluctuating from an F1 score of 0.944 with 20 windows to 0.901 with 28 windows, demonstrating its less stable performance compared to random sampling and highlighting its difficulty in finding diverse yet representative windows for adaptation. Similarly, the time-based method, constrained by its lack of diversity, consistently results in lower F1 scores across different training window sizes.

\section{\noindent\textbf{F1 SCORE ACROSS VARYING SAMPLING RATES AND WINDOW TIME SPANS}}
\label{appendix:vary_sample_rate_window_size}
        
\begin{table}[htbb!]
    \centering
    \caption{F1 scores across different sampling rates and window time spans for User Identification.}
    \label{tab:mod1f1}
    \begin{tabular}{|c|c|c|c|c|c|c|}
    \hline
    \multicolumn{2}{|c|}{\textbf{}} & \multicolumn{5}{c|}{\textbf{Window time span}} \\
    \hline
    \multicolumn{2}{|c|}{\textbf{}} & \textbf{1~s} & \textbf{2~s} & \textbf{5~s} & \textbf{10~s} & \textbf{20~s} \\
    \hline
    \multirow{5}{*}{\textbf{Frequency}} 
    & \textbf{1~Hz}  & \smallpm{0.842}{0.153} & \smallpm{0.893}{0.125} & \smallpm{0.952}{0.088} & \smallpm{0.925}{0.090} & \smallpm{0.912}{0.0076} \\
    \cline{2-7}
    & \textbf{2~Hz}  & \smallpm{0.891}{0.170} & \smallpm{0.911}{0.122} & \smallpm{0.980}{0.028} & \smallpm{0.963}{0.057} & \smallpm{0.942}{0.181} \\
    \cline{2-7}
    & \textbf{5~Hz}  & \smallpm{0.959}{0.127} & \smallpm{0.990}{0.015} & \smallpm{0.971}{0.026} & \smallpm{0.979}{0.040} & \smallpm{0.980}{0.032} \\
    \cline{2-7}
    & \textbf{10~Hz} & \smallpm{0.987}{0.031} & \smallpm{0.964}{0.102} & \smallpm{0.947}{0.102} & \smallpm{0.976}{0.030} & \smallpm{0.966}{0.063} \\
    \cline{2-7}
    & \textbf{20~Hz} & \smallpm{0.991}{0.009} & \smallpm{0.995}{0.005} & \smallpm{0.991}{0.009} & \smallpm{0.991}{0.007} & \smallpm{0.983}{0.019} \\
    \hline
    \end{tabular}
\end{table}

     We trained, fine-tuned, and evaluated our owner identification module with different data sampling rates and input window sizes. Table~\ref{tab:mod1f1} shows the F1 score for a variety of configurations, which we found is highest (F1: 0.995) using a sampling rate of 20 Hz and a window duration of 2 seconds. Across all configurations, even with lower sampling rates and shorter window sizes, the F1 scores remain consistently high (\texttt{>}~0.84), demonstrating robustness and requiring less computation to process. 
    
    Interestingly, we observe a plateau or slight performance drop with certain configurations, such as a 5 Hz sampling rate and a 5-second window size. We hypothesize that while larger window sizes and higher sampling rates provide more information, they require a larger and more expressive model to generalize well. Our model has the smallest number of parameters, compared to existing authentication works, which limits its capability of generalizing to higher dimensional inputs. Despite these limitations, our results suggest that a highly expressive model is not strictly necessary. The best configuration performs admirably even at moderately low sampling rates, making the system computationally efficient and power-friendly. Additionally, the stability of F1 scores across varying configurations highlights the model's adaptability, offering flexibility in deployment scenarios where power and processing constraints are critical.

\section{EVALUATING THE OPTIMAL NUMBER OF LSTM HEADS}
\label{appendix:num_lstm_heads}

\begin{table}[htbp!]
\centering
\caption{Varying the number of LSTM heads and output embedding size}
\label{tab:lstm-heads}
\begin{tabular}{cccc}
\toprule
\textbf{Configurations} & \textbf{Partners} & \textbf{Strangers} & \textbf{Overall} \\
\midrule
1H & \smallpm{0.672}{0.178} & \smallpm{0.794}{0.125} & \smallpm{0.733}{0.165} \\
2H & \smallpm{0.743}{0.141} & \smallpm{0.807}{0.105} & \smallpm{0.775}{0.128} \\
4H & \smallpm{0.928}{0.061} & \smallpm{0.927}{0.090} & \smallpm{0.928}{0.077} \\
\textbf{8H} & \textbf{\smallpm{0.961}{0.025}} & \textbf{\smallpm{0.948}{0.039}} & \textbf{\smallpm{0.954}{0.033}} \\
12H & \smallpm{0.934}{0.043} & \smallpm{0.833}{0.207} & \smallpm{0.883}{0.158} \\
16H & \smallpm{0.932}{0.057} & \smallpm{0.915}{0.049} & \smallpm{0.924}{0.054} \\
1H\_EMB32 & \smallpm{0.952}{0.030} & \smallpm{0.898}{0.091} & \smallpm{0.925}{0.073} \\
\bottomrule
\end{tabular}
\end{table}

        We varied the number of LSTM heads to assess their impact on identity recognition. As shown in Table~\ref{tab:lstm-heads}, performance peaks at 8 heads (F1 = 0.954); beyond that, additional heads offer diminishing returns or lead to slight degradation, confirming the effectiveness of our chosen configuration.

\section{USER IDENTIFICATION CLASSIFICATION HEAD CHOICE}
\label{appendix:id_classification_head_choice}

\begin{figure}[htbp!]
\centering\includegraphics[width=0.6\linewidth]{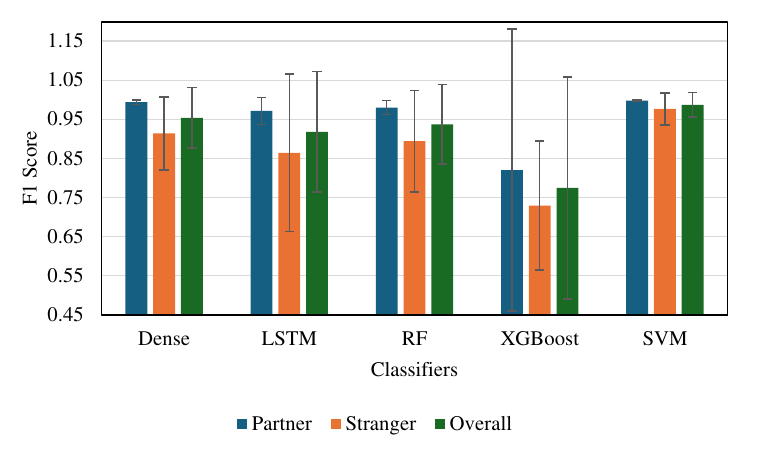}%
\caption{Exploring different choices of user identity classifiers.}\label{fig:mod1clf}
\end{figure}

     We evaluate a range of classifiers to assess their performance for user identification. For a comprehensive analysis, we include traditional machine learning approaches, such as Support Vector Machine (SVM) and Random Forest (RF), as well as deep learning-based methods like Long Short-Term Memory (LSTM) and Dense Neural Networks. Additionally, we include XGBoost, a widely used ensemble learning algorithm. Specifically, we set the SVM with an RBF kernel, Random Forest with 100 estimators, and use a multi-head LSTM (each with 6 units) to mirror the AutoEncoder architecture. For the dense network, we adopt two dense layers connected to the AutoEncoder output for prediction. The XGBoost model is configured with a learning rate of 0.1 (commonly used to balance convergence speed and performance), a maximum tree depth of 6, 100 estimators, and the logloss evaluation metric to optimize classification performance.
    
    Figure~\ref{fig:mod1clf} shows the performance of \sys with different backbones for the classifier. We adopt SVM as the backbone of the classifier, since it has the highest performance, with an F1 score of 0.98. Among the deep learning-based models, the Dense classifier demonstrates an F1 score of 0.969, which is slightly lower than SVM but higher than the LSTM classifier. Out of all architectures, SVM is the least complex and least expressive, but is less susceptible to overfitting. Its high performance suggests that the encoder could extract highly relevant features, with little data, to distinguish between the phone owner and other users, which is helpful in our few-shot fine-tuning process. Other more expressive methods, such as Dense MLPs and LSTM, typically require more data, memory, and compute resources.

\section{GENERALIZATION TO UNSEEN APPS}
\label{appendix:app_held_out}

\begin{table}[]
\centering
\caption{Accuracy of \sys~under app-held-out training. Accuracy drops are shown in parentheses}
\label{tab:app_held_out}
\begin{tabular}{lccc}
\hline
\textbf{Accuracy} & \textbf{Category (5 classes)} & \textbf{Action (9 classes)} & \textbf{Subaction (28 classes)} \\
\hline
Top-1 & \smallpm{0.6384}{0.0710}  \scriptsize{(0.0134)} & \smallpm{0.5885}{0.0805}  \scriptsize{(0.0075)} & \smallpm{0.4964}{0.0963}  \scriptsize{(0.0696)} \\
Top-2 & \smallpm{0.8555}{0.0562}  \scriptsize{(0.0294)} & \smallpm{0.7734}{0.0795}  \scriptsize{(0.0350)} & \smallpm{0.6527}{0.0987}  \scriptsize{(0.0560)} \\
Top-3 & \smallpm{0.9455}{0.0274}  \scriptsize{(0.0131)} & \smallpm{0.8742}{0.0629}  \scriptsize{(0.0259)} & \smallpm{0.7404}{0.0958}  \scriptsize{(0.0545)} \\
Top-5 & --                                              & \smallpm{0.9648}{0.0286}  \scriptsize{(0.0088)} & \smallpm{0.8557}{0.0777}  \scriptsize{(0.0357)} \\
\hline
\end{tabular}
\end{table}

We simulated the case where an app is unseen during training but present in deployment: for each of the six apps, we remove all of its samples from the training split while leaving the test set unchanged. This yields six folds; each model is trained on five apps and evaluated on the full six-app test set (Table \ref{tab:app_held_out}). We observed minor drops in accuracies across different levels and top-$k$s, with the average, median, and maximum drop of 3.6\%, 2.9\%, and 7.0\% respectively, compared to the standard seen-app training baseline (Figure \ref{fig:mod2perf}). The small degradation suggests that \sys~generalizes well to apps it never observed, supporting its further use on larger, more diverse app populations.

\section{FULL ACTION LIST}
\label{appendix:full_action_list}
Table~\ref{tab:action_list} lists all the tasks participants are required to complete. Only the Platform and Subaction columns are visible to participants.

\begin{table}[]
    \centering\caption{Full Action List.}
        \label{tab:action_list}
\begin{adjustbox}{width=\linewidth}     
\begin{tabular}{ccccc}
\toprule
\textbf{Task ID} &
  \textbf{Platform} &
  \textbf{\begin{tabular}[c]{@{}c@{}}Category \\ (Invisible to participant)\end{tabular}} &
  \textbf{\begin{tabular}[c]{@{}c@{}}Action \\ (Invisible to participant)\end{tabular}} &
  \textbf{\begin{tabular}[c]{@{}c@{}}Subaction \\ (Visible to participant)\end{tabular}} \\
  \midrule
1  & Gmail     & ACCESS       & View content           & View emails                                   \\
2  & Gmail     & ACCESS       & View account           & View account settings                         \\
3  & Spotify   & ACCESS       & View account           & View account settings                         \\
4  & Spotify   & ACCESS       & View account           & Subscription details                          \\
5  & Amazon    & ACCESS       & View account           & Inspect order history                         \\
6  & Amazon    & ACCESS       & View account           & View browsing history                         \\
7  & Amazon    & ACCESS       & View account           & View payment settings                         \\
8  & Instagram & ACCESS       & View content           & See your account's post history               \\
9  & Instagram & ACCESS       & Upload content         & Post a photograph                             \\
10 & Instagram & ACCESS       & View account           & Inspect account info                          \\
11 & YouTube   & ACCESS       & View content           & View Watch history                            \\
12 & YouTube   & ACCESS       & Upload content         & Upload video                                  \\
13 & YouTube   & ACCESS       & View account           & Open account settings                         \\
14 & Slack     & ACCESS       & View content           & View messages                                 \\
15 & Slack     & ACCESS       & View content           & Inspect files                                 \\
16 & Slack     & ACCESS       & View account           & Open account settings                         \\
17 & Gmail     & Modification & Alter account settings & Change profile photo                          \\
18 & Gmail     & Modification & Modify content         & Delete emails                                 \\
19 & Spotify   & Modification & Alter account settings & Change profile photo                          \\
20 &
  Spotify &
  Modification &
  Alter account settings &
  \begin{tabular}[c]{@{}c@{}}Try to change account email \\ to ilove2025@gmail.com\end{tabular} \\
21 & Spotify   & Modification & Alter account settings & Change account name                           \\
22 & Spotify   & Modification & Modify content         & Add, edit, delete music list                  \\
23 & Amazon    & Modification & Alter account settings & Change password (guess current password)      \\
24 & Amazon    & Modification & Alter account settings & Change address                                \\
25 & Instagram & Modification & Alter account settings & Change username                               \\
26 & Instagram & Modification & Alter account settings & Change profile picture                        \\
27 & YouTube   & Modification & Alter account settings & Change password (ask for current password)    \\
28 & Slack     & Modification & Alter files            & Upload a file                                 \\
29 & Slack     & Modification & Alter files            & Modify/rename/comment on a file in group chat \\
30 & Slack     & Modification & Alter files            & Delete a file uploaded by your account        \\
31 & Slack     & Modification & Alter account settings & Change password (guess current password)      \\
32 & Gmail     & POST         & Send content           & Send emails                                   \\
33 &
  Amazon &
  POST &
  Send content &
  \begin{tabular}[c]{@{}c@{}}Send product reviews \\ (neutral/decent comments)\end{tabular} \\
34 & Instagram & POST         & Send content           & Send direct message to a follower             \\
35 & YouTube   & POST         & Send content           & Comment (neutral/decent comments)             \\
36 & Slack     & POST         & Send content           & Send message                                  \\
37 &
  File &
  SOFTWARE &
  Install SOFTWARE &
  \begin{tabular}[c]{@{}c@{}}Install SOFTWARE \\ (test.apk, located in Files app)\end{tabular} \\
38 & Spotify   & GENERAL      & GENERAL                & Music listening for some time                 \\
39 & Amazon    & GENERAL      & GENERAL                & Search items                                  \\
40 & Amazon    & GENERAL      & GENERAL                & Browse item info for some time                \\
41 & Instagram & GENERAL      & GENERAL                & Read others' posts for some time              \\
42 & Instagram & GENERAL      & GENERAL                & Like posts                                    \\
43 & YouTube   & GENERAL      & GENERAL                & Search for others' videos                     \\
44 & YouTube   & GENERAL      & GENERAL                & Watch for some time                           \\
\bottomrule
\end{tabular}
\end{adjustbox}
\end{table}

\section{\LONG{SENSITIVITY TO NON-OWNER CONTAMINATION IN RECALIBRATION}}
\label{app:contamination}
     
\LONG{We further evaluate session-level recalibration under a controlled contamination-and-recovery setting. Each case contains four longitudinal sessions. Session~1 is used for clean initialization, and one intermediate session, either Session~2 or Session~3, is contaminated by adding the first \(x\) minutes of non-owner data to the first three minutes of owner calibration data. The non-owner samples are intentionally treated as owner calibration data, yielding contamination ratios of \(x/(3+x)\). After the contaminated session, later sessions resume clean owner-data recalibration. We report end-to-end Top-1 event-flagging F1 over the full four-session sequence, averaged over four cases from two participant pairs.}

\LONG{As shown in Table~\ref{tab:contam_recover_horizontal}, performance decreases as the contamination ratio increases, from \(0.865 \pm 0.039\) with no contamination to \(0.794 \pm 0.022\) at 33.3\% contamination. This result indicates that contaminated recalibration data can affect AID, but the effect is not necessarily persistent when later calibration windows again contain clean owner data. Since the owner is expected to generate the majority of phone-use traces in our target setting, this controlled experiment suggests that an isolated contaminated recalibration step can be partially corrected by subsequent owner-data recalibration, allowing AID to retain relatively stable longitudinal performance.}

\begin{table}[t]
\centering
\caption{Sensitivity of longitudinal performance to one contaminated recalibration event. Each case evaluates all four longitudinal sessions. One intermediate session's calibration buffer is contaminated with non-owner samples, while later sessions resume clean owner recalibration. Results report end-to-end Top-1 event-flagging F1, averaged over four contamination cases from two participant pairs.}
\label{tab:contam_recover_horizontal}
\begin{tabular}{lcccccc}
\toprule
Contam. ratio & 0.0\% & 9.1\% & 16.7\% & 23.1\% & 28.6\% & 33.3\% \\
\midrule
E2E Top-1 F1 & 
$0.865 \pm 0.039$ &
$0.836 \pm 0.022$ &
$0.808 \pm 0.016$ &
$0.802 \pm 0.042$ &
$0.802 \pm 0.032$ &
$0.794 \pm 0.022$ \\
\bottomrule
\end{tabular}
\end{table}


\end{document}